\documentclass[aip,reprint,amsmath,amssymb,superscriptaddress]{revtex4-1}

\usepackage{verbatim} 

\usepackage{graphicx}
\usepackage{dcolumn}

\usepackage{amstext,amsmath,amssymb,amsfonts}
\usepackage{bm}

\usepackage{txfonts}
\usepackage{mathtools}
\DeclareMathAlphabet{\mathpzc}{OT1}{pzc}{m}{it}
\DeclareFontFamily{OT1}{pzc}{}
\DeclareFontShape{OT1}{pzc}{m}{it}{<-> s * [1.100] pzcmi7t}{}
\DeclareMathAlphabet{\mathpzc}{OT1}{pzc}{m}{it}

\usepackage{hyperref}

\usepackage{color}
\definecolor{lightblue}{rgb}{0.2,0.2,0.7}
\definecolor{darkblue}{rgb}{0,0.25,0.5}
\definecolor{redbrown}{rgb}{0.875,0.25,0.125}
\definecolor{darkgreen}{rgb}{0,0.5,0}

\renewcommand{\b}[1]{\ensuremath{\mathbf{#1}}}
\renewcommand{\H}{\ensuremath{\text{H}}}

\renewcommand{\l}{\ensuremath{\lambda}}

\newcommand{\lr}{\ensuremath{\text{lr}}}
\newcommand{\sr}{\ensuremath{\text{sr}}}
\newcommand{\ee}{\ensuremath{\text{ee}}}
\renewcommand{\ne}{\ensuremath{\text{ne}}}

\newcommand{\HF}{\ensuremath{\text{HF}}}

\renewcommand{\d}{\ensuremath{\text{d}}}
\newcommand{\s}{\ensuremath{\text{s}}}

\newcommand{\x}{\ensuremath{\text{x}}}
\newcommand{\xc}{\ensuremath{\text{xc}}}

\newcommand{\Hxc}{\ensuremath{\text{Hxc}}}

\DeclareMathOperator{\erf}{erf}

\newcommand{\isEquivTo}[1]{\underset{#1}{\sim}}

\renewcommand{\i}{\ensuremath{\text{i}}}

\begin{document}

\title{Photoionization and core resonances from range-separated density-functional theory: General formalism and example of the beryllium atom}

\author{Karno Schwinn}
\email{karno.schwinn@lct.jussieu.fr}
\affiliation{Laboratoire de Chimie Th\'eorique, Sorbonne Universit\'e and CNRS, F-75005 Paris, France}
\author{Felipe Zapata}
\email{felipe.zapata@matfys.lth.se}
\affiliation{Department of Physics, Lund University, Box 118, SE-221 00 Lund, Sweden}
\author{Antoine Levitt}
\email{antoine.levitt@inria.fr}
\affiliation{CERMICS, \'Ecole des Ponts and Inria Paris, 6 \& 8 Avenue Blaise Pascal, 77455 Marne-la-Vall\'ee, France}
\author{\'Eric Canc\`es}
\email{eric.cances@enpc.fr}
\affiliation{CERMICS, \'Ecole des Ponts and Inria Paris, 6 \& 8 Avenue Blaise Pascal, 77455 Marne-la-Vall\'ee, France}
\author{Eleonora Luppi}
\email{eleonora.luppi@lct.jussieu.fr}
\affiliation{Laboratoire de Chimie Th\'eorique, Sorbonne Universit\'e and CNRS, F-75005 Paris, France}
\author{Julien Toulouse}
\email{toulouse@lct.jussieu.fr}
\affiliation{Laboratoire de Chimie Th\'eorique, Sorbonne Universit\'e and CNRS, F-75005 Paris, France}
\affiliation{Institut Universitaire de France, F-75005 Paris, France}

\date{March 11, 2022}

\begin{abstract}
We explore the merits of linear-response range-separated time-dependent density-functional theory (TDDFT) for the calculation of photoionization spectra. We consider two variants of range-separated TDDFT, namely the time-dependent range-separated hybrid (TDRSH) scheme  which uses a global range-separation parameter and the time-dependent locally range-separated hybrid (TDLRSH) which uses a local range-separation parameter, and compare with standard time-dependent local-density approximation (TDLDA) and time-dependent Hartree-Fock (TDHF). We show how to calculate photoionization spectra with these methods using the Sternheimer approach formulated in a non-orthogonal B-spline basis set with appropriate frequency-dependent boundary conditions. We illustrate these methods on the photoionization spectrum of the Be atom, focusing in particular on the core resonances. Both the TDRSH and TDLRSH photoionization spectra are found to constitute a large improvement over the TDLDA photoionization spectrum and a more modest improvement over the TDHF photoionization spectrum.
\end{abstract}

\maketitle

\section{Introduction}

Time-dependent density-functional theory (TDDFT)~\cite{RunGro-PRL-84}, applied within the adiabatic linear-response formalism~\cite{GroKoh-PRL-85,Cas-INC-95,PetGosGro-PRL-96}, is a widely used approach for calculating bound-state excitations in electronic systems. Less commonly, linear-response TDDFT has also been used for calculating photoionization spectra (electronic transitions from bound to continuum states) of atoms and molecules~\cite{ZanSov-PRA-80,LevSov-PRA-84,SteDecLis-JPB-95,SteAltFroDec-CP-97,SteDec-JPB-97,SteDec-JCP-00,SteDecGor-JCP-01,NakYab-JCP-01,WasMaiBur-PRL-03,SteFroDec-JCP-05,SteTofFroDec-JCP-06,TofSteDec-PRA-06,YabNakIwaBer-PSS-06,SteTofFroDec-TCA-07,ZhoChu-PRA-09,YanFaaBur-JCP-09}. These calculations require an appropriate description of the continuum states (e.g., using grid-based approaches or B-spline basis sets) and an accurate enough exchange-correlation potential and associated response kernel. Semilocal density-functional approximations, such as the local-density approximation (LDA) or generalized-gradient approximations (GGA), do not usually provide accurate atomic and molecular photoionization spectra. These approximations suffer indeed from large self-interaction errors and exponentially decaying exchange-correlation potentials, leading to too low ionization thresholds and resonances that are either at too low energies or completely absent. More satisfactory photoionization spectra are obtained with asymptotically corrected exchange-correlation potential approximations~\cite{SteDecLis-JPB-95,SteAltFroDec-CP-97,SteDec-JPB-97,SteDec-JCP-00,SteFroDec-JCP-05,SteTofFroDec-JCP-06,TofSteDec-PRA-06,SteTofFroDec-TCA-07} (restoring the correct $-1/r$ long-range asymptotic decay) and with the more involved exact-exchange (EXX) potential~\cite{SteDecGor-JCP-01} or the localized Hartree-Fock (HF) exchange potential and its associated kernel~\cite{ZhoChu-PRA-09}.

An alternative for overcoming the limitations of TDDFT with semilocal density-functional approximations is given by range-separated TDDFT approaches~\cite{TawTsuYanYanHir-JCP-04,YanTewHan-CPL-04,PeaHelSalKeaLutTozHan-PCCP-06,LivBae-PCCP-07,BaeLivSal-ARPC-10,TsuSonSuzHir-JCP-10,FroKneJen-JCP-13,RebSavTou-MP-13} which, in the simplest variant, express the long-range part of the exchange potential and kernel at the HF level while a semilocal density-functional approximation is still used for the short-range part of the kernel. Range-separated TDDFT appropriately describes Rydberg and charge-transfer electronic excitations, and has become very much used in calculations of bound-state excitations in molecules. In Ref.~\onlinecite{ZapLupTou-JCP-19}, some of the present authors started to explore the merits of range-separated TDDFT for the calculation of photoionization spectra and showed that the so-called linear-response time-dependent range-separated hybrid (TDRSH) scheme~\cite{RebSavTou-MP-13,TouRebGouDobSeaAng-JCP-13} provides an adequate photoionization spectrum of the He atom. In the present work, we continue the systematic exploration of linear-response TDRSH for the calculation of photoionization spectra. We focus on the Be atom which has a much richer photoionization spectrum than the He atom since it contains both core and valence electrons, leading in particular to a series of core resonances (1s$\to$2p, 1s$\to$3p, etc...) just below the 1s ionization edge. We also test a new variant, called the linear-response time-dependent locally range-separated hybrid (TDLRSH) scheme, in which the range-separation parameter is a position-dependent function~\cite{KruScuPerSav-JCP-08,AscKum-JCP-19,KlaBah-JCTC-20,MaiIkaNak-JCP-21}, which allows for more flexibility in the description of both valence and core properties. 

As in Ref.~\onlinecite{ZapLupTou-JCP-19}, we use a B-spline basis set
for an appropriate description of the continuum. More specifically, in
Ref.~\onlinecite{ZapLupTou-JCP-19}, we used a straightforward
diagonalization of the linear-response Casida equations (in the
orthogonal occupied/virtual orbital basis) using zero boundary
conditions at the edge of the support of the last B-spline function,
resulting in a discretization of the continuous spectrum. Whereas in
the case of the He atom the relatively simple structure of the
photoionization spectrum (only one channel) made it possible to use a simple interpolation scheme for the oscillator strengths, this approach is not feasible for the more complicated photoionization spectrum of the Be atom and one would have to use an artificial broadening to compensate. In this work, we use instead the linear-response Sternheimer approach~\cite{MahSub-BOOK-90,SteDecLis-JPB-95,SteDec-JCP-00,YabNakIwaBer-PSS-06,AndBotMarRub-JCP-07,StrLehRubMarLou-INC-12,HofSchKum-JCP-18,HofSchKum-PRA-19,HofKum-JCP-20} (in the non-orthogonal B-spline basis) using appropriate frequency-dependent boundary conditions, resulting directly in an adequate representation of the continuous spectrum without the need for broadening. This is a much more efficient way of calculating photoionization spectra over a wide energy window, requiring only a relatively small computational box. 

The paper is organized as follows. In Section~\ref{sec:theory}, we review the range-separated hybrid (RSH) and locally range-separated hybrid (LRSH) schemes, and give in some details the linear-response Sternheimer equations including a nonlocal HF exchange kernel both in real space and in a general non-orthogonal basis set which, to the best of our knowledge, were never given in the literature. We also discuss how the boundary conditions on a finite domain are imposed and our specific implementation using a B-spline basis set. In Section~\ref{sec:results}, we give and discuss the results obtained on the Be atom. We explain how to select an optimal range-separation parameter, we discuss the photoionization spectra at the TDRSH and TDLRSH level and compare with linear-response time-dependent local-density approximation (TDLDA) and time-dependent Hartree-Fock (TDHF), and we analyze the positions and the Fano lineshape of the core resonances. Finally, Section~\ref{sec:conclusion} contains our conclusions. In the Appendix, we explain in details how to obtain the appropriate boundary conditions for atoms.

\section{Theory and computational method}
\label{sec:theory}

For simplicity, we consider only the case of a closed-shell atomic or molecular system and thus work on the spin-free one-electron Hilbert space $L^2(\mathbb{R}^3,\mathbb{C})$. Unless otherwise indicated, Hartree atomic units are used in this work.

\subsection{Range-separated hybrid scheme}

Let us briefly recall the range-separated hybrid (RSH) scheme~\cite{AngGerSavTou-PRA-05}. The RSH orbitals $\{\varphi_i\}$ and their associated energies $\{\varepsilon_i\}$ of a $N$-electron system are found from the self-consistent Schr\"odinger-type equation
\begin{eqnarray}
\int_{\mathbb{R}^3} h[\gamma_0](\b{r},\b{r}') \varphi_i(\b{r}') \d \b{r}' = \varepsilon_i \varphi_i(\b{r}),
\label{}
\end{eqnarray}
where $h[\gamma_0](\b{r},\b{r}')$ is the nonlocal RSH Hamiltonian depending on the density matrix $\gamma_0(\b{r},\b{r}') = 2\sum_{i=1}^{N/2} \varphi_i(\b{r}) \varphi_i^*(\b{r}')$. The RSH Hamiltonian has the form, for a generic density matrix $\gamma$,
\begin{eqnarray}
h[\gamma](\b{r},\b{r}') = T(\b{r},\b{r}') + \delta(\b{r}-\b{r}') v_{\text{ne}}(\b{r}) + v_\Hxc[\gamma](\b{r},\b{r}'), \;
\label{}
\end{eqnarray}
where $T(\b{r},\b{r}')$ is the kinetic integral kernel such that $\int_{\mathbb{R}^3} T(\b{r},\b{r}') \varphi_i(\b{r}') \d \b{r}' = -(1/2) \bm{\nabla}^2 \varphi_i(\b{r})$, and $v_{\text{ne}}(\b{r})$ is the nuclei-electron potential and $v_\Hxc[\gamma](\b{r},\b{r}')$ is the Hartree-exchange-correlation potential. The expression of $v_\Hxc[\gamma](\b{r},\b{r}')$ is
\begin{eqnarray}
v_\Hxc[\gamma](\b{r},\b{r}') &=& \delta(\b{r}-\b{r}') v_\H[\rho_\gamma](\b{r}) + v_\x^{\lr,\HF}[\gamma](\b{r},\b{r}') 
\nonumber\\
&& + \delta(\b{r}-\b{r}') v_\xc^{\sr}[\rho_\gamma](\b{r}),
\label{vHxc}
\end{eqnarray}
containing the local Hartree potential 
\begin{eqnarray}
v_\text{H}[\rho_\gamma](\b{r}) = \int_{\mathbb{R}^3} \rho_\gamma(\b{r}') w_\ee(\b{r},\b{r}') \d \b{r}',
\end{eqnarray}
written with the density $\rho_\gamma(\b{r})=\gamma(\b{r},\b{r})$ and the Coulomb electron-electron interaction $w_\ee(\b{r},\b{r}')=1/|\b{r}-\b{r}'|$, the nonlocal long-range (lr) HF exchange potential 
\begin{eqnarray}
v_{\text{x}}^{\lr,\HF}[\gamma](\b{r},\b{r}') = - \frac{1}{2}\gamma(\b{r},\b{r}') w_\ee^\lr(\b{r},\b{r}'),
\label{vxlrHF}
\end{eqnarray}
written with the long-range electron-electron interaction~\cite{Sav-INC-96} 
\begin{eqnarray}
w_\ee^\lr(\b{r},\b{r}')=\frac{\erf(\mu|\b{r}-\b{r}'|)}{|\b{r}-\b{r}'|},
\label{weelrerf}
\end{eqnarray}
with $\mu = \tilde{\mu}/a_0$ where $a_0=1$ a.u. is the Bohr radius and $\tilde{\mu}\in [0,+\infty)$ is the adimensional range-separation parameter, and the local complementary short-range (sr) exchange-correlation potential $v_\xc^{\sr}[\rho_\gamma](\b{r})$. For the latter term, we use in this work the LDA
\begin{eqnarray}
v_\xc^{\sr}[\rho_\gamma](\b{r}) = \left. \frac{\partial \bar{e}_{\xc,\text{UEG}}^\sr(\rho,\mu)}{\partial \rho} \right|_{\rho=\rho_\gamma(\b{r})},
\label{vxcsr}
\end{eqnarray}
where $\bar{e}_{\xc,\text{UEG}}^\sr(\rho,\mu)$ is the complementary short-range exchange-correlation energy density of the uniform-electron gas (UEG), as parametrized in Ref.~\onlinecite{PazMorGorBac-PRB-06}.

For more flexibility in the description of both valence and core properties, we will also consider an extension of the RSH scheme, referred to as the locally range-separated hybrid (LRSH) scheme, in which the range-separation parameter $\mu$ in Eqs.~(\ref{weelrerf}) and~(\ref{vxcsr}) is replaced by a function of position $\b{r} \mapsto \mu(\b{r})$ (see Refs.~\onlinecite{KruScuPerSav-JCP-08,AscKum-JCP-19,KlaBah-JCTC-20,MaiIkaNak-JCP-21}). The long-range electron-electron interaction in Eq.~(\ref{weelrerf}) now becomes~\cite{KlaBah-JCTC-20}
\begin{eqnarray}
w_\ee^\lr(\b{r},\b{r}')=\frac{1}{2} \left[ \frac{\erf(\mu(\b{r})|\b{r}-\b{r}'|)}{|\b{r}-\b{r}'|} + \frac{\erf(\mu(\b{r}')|\b{r}-\b{r}'|)}{|\b{r}-\b{r}'|}\right].
\label{weelrerfmur}
\end{eqnarray}
Following Ref.~\onlinecite{KruScuPerSav-JCP-08}, we choose $\mu(\b{r})$ as
\begin{eqnarray}
\mu(\b{r}) = \frac{\tilde{\mu}}{2} \frac{|\nabla \rho(\b{r})|}{\rho(\b{r})},
\end{eqnarray}
where again $\tilde{\mu} \in [0,+\infty)$ is the adimensional range-separation parameter and we take $\rho(\b{r})$ as the fixed HF density. For the hydrogen atom, $\rho(\b{r}) \propto e^{-2|\b{r}|/a_0}$, and thus $\mu(\b{r}) = \tilde{\mu}/a_0$, i.e. the LRSH scheme reduces to the RSH scheme.

For $\tilde{\mu}=0$, the long-range interaction $w_\ee^\lr$ vanishes and $v_\xc^{\sr}$ becomes the usual full-range LDA exchange-correlation potential, and thus the RSH and LRSH schemes reduce to standard Kohn-Sham (KS) LDA. For $\tilde{\mu}\to\infty$, the long-range interaction $w_\ee^\lr$ becomes the usual Coulomb interaction and $v_\text{xc}^{\sr}$ vanishes, and thus the RSH and LRSH schemes reduce to standard HF. Typically, in between these two limits, one expect to find an intermediate value of $\tilde{\mu}$ leading to properties more accurate than those given by either KS LDA or HF.

\subsection{Linear-response Sternheimer equations in real space}

We now formulate linear response of the RSH or LRSH scheme using the Sternheimer approach. Even though the Sternheimer approach is well known for TDDFT without nonlocal HF exchange~\cite{MahSub-BOOK-90,SteDecLis-JPB-95,SteDec-JCP-00,YabNakIwaBer-PSS-06,AndBotMarRub-JCP-07,StrLehRubMarLou-INC-12,HofSchKum-JCP-18,HofSchKum-PRA-19,HofKum-JCP-20}, we did not find real-space expressions for the case including nonlocal HF exchange in the literature.

We consider the following time-dependent perturbation potential
\begin{eqnarray}
v_\text{ext}(\b{r},t) = \left[ v_\text{ext}(\b{r}) e^{-\i \omega t} + v_\text{ext}(\b{r}) e^{+\i \omega t} \right] e^{\eta t},
\label{vextt}
\end{eqnarray}
where $v_\text{ext}(\b{r}) = \b{r} \, \cdot \, {\cal E} \, \b{e}$ is the electric-dipole interaction (${\cal E}$ is the amplitude of the electric field and $\b{e}$ is its unit polarization vector), $\omega \geq 0$ is the frequency, and $e^{\eta t}$ is an adiabatic switching factor with a small parameter $\eta > 0$ so that $v_\text{ext}(\b{r},t\to -\infty) = 0$ (see Refs.~\onlinecite{HofSchKum-JCP-18,DupLev-ARX-21} for a discussion about the parameter $\eta$). The time-dependent occupied RSH orbitals $\{ \psi_i \}$ satisfy the time-dependent Schr\"odinger-type equation 
\begin{eqnarray}
 \i \frac{\partial}{\partial t} \psi_i(\b{r},t) = \int_{\mathbb{R}^3} h[\gamma(t)](\b{r},\b{r}') \psi_i(\b{r}',t) \d \b{r}' + v_{\text{ext}}(\b{r},t)  \psi_i(\b{r},t),
\nonumber\\
\label{TDRSH}
\end{eqnarray}
where the RSH Hamiltonian is now evaluated at the time-dependent density matrix $\gamma(\b{r},\b{r}',t) = 2\sum_{i=1}^{N/2} \psi_i(\b{r},t) \psi_i^*(\b{r}',t)$. We expand the time-dependent RSH occupied orbitals to first order in the electric field ${\cal E}$ as
\begin{eqnarray}
\psi_i(\b{r},t) = \left( \varphi_i(\b{r}) + {\cal E} \; \psi_i^{(1)}(\b{r},t)  \right) e^{-\i \varepsilon_i t} + O({\cal E}^2),
\label{psiit}
\end{eqnarray}
where $\varphi_i \equiv \psi_i^{(0)}$ are the zeroth-order (time-independent) orbitals. Inserting Eq.~(\ref{psiit}) into Eq.~(\ref{TDRSH}), and keeping only first-order terms, leads to the equation for $\psi_i^{(1)}$
\begin{eqnarray}
&&\left( \i \frac{\partial}{\partial t} +\varepsilon_i \right) \psi_i^{(1)}(\b{r},t) = \int_{\mathbb{R}^3} h[\gamma_0](\b{r},\b{r}') \psi_i^{(1)}(\b{r}',t) \d\b{r}'
\nonumber\\
&& +  \int_{\mathbb{R}^3} v_\Hxc^{(1)}(\b{r},\b{r}',t) \varphi_i(\b{r}') \d\b{r}' + v_\text{ext}^{(1)}(\b{r},t) \varphi_i(\b{r}),
\label{eqforpsii1}
\end{eqnarray}
where we have introduced $v_\text{ext}^{(1)}(\b{r},t) = v_\text{ext}(\b{r},t)/{\cal E}$ and the first-order change in the Hartree-exchange-correlation potential
\begin{eqnarray}
v_\Hxc^{(1)}(\b{r}_1,\b{r}_1',t) = \;\;\;\;\;\;\;\;\;\;\;\;\;\;\;\;\;\;\;\;\;\;\;\;\;\;\;\;\;\;\;\;\;\;\;\;\;\;\;\;\;\;\;\;\;\;\;\;\;\;\;\;
\nonumber\\
 \int_{\mathbb{R}^6} f_\Hxc[\gamma_0](\b{r}_1,\b{r}_1';\b{r}_2,\b{r}_2') \gamma^{(1)}(\b{r}_2,\b{r}_2',t) \d\b{r}_2 \d\b{r}_2',
\label{}
\end{eqnarray}
involving the first-order change in the density matrix
\begin{eqnarray}
\gamma^{(1)}(\b{r},\b{r}',t) = 2\sum_{i=1}^{N/2} \left[ \psi_i^{(1)}(\b{r},t) \varphi_i^*(\b{r}') + \varphi_i(\b{r}) \psi_i^{(1)*}(\b{r}',t) \right],
\label{}
\end{eqnarray}
and the Hartree-exchange-correlation kernel
\begin{eqnarray}
f_\Hxc[\gamma_0](\b{r}_1,\b{r}_1';\b{r}_2,\b{r}_2') = \left. \frac{\delta v_\Hxc[\gamma](\b{r}_1,\b{r}_1')}{\delta \gamma(\b{r}_2,\b{r}_2')} \right|_{\gamma=\gamma_0}.
\label{}
\end{eqnarray}
From Eq.~(\ref{vHxc}), the latter quantity is found to be
\begin{eqnarray}
f_\Hxc[\gamma_0](\b{r}_1,\b{r}_1';\b{r}_2,\b{r}_2') = \delta(\b{r}_1-\b{r}_1') \delta(\b{r}_2-\b{r}_2') f_\H(\b{r}_1,\b{r}_2) \;\;\;\;
\nonumber\\
+ f_\H^{\lr,\HF}(\b{r}_1,\b{r}_1';\b{r}_2,\b{r}_2')
+ \delta(\b{r}_1-\b{r}_1') \delta(\b{r}_2-\b{r}_2') f_\xc^{\sr}[\rho_{\gamma_0}](\b{r}_1,\b{r}_2),
\nonumber\\
\label{}
\end{eqnarray}
where $f_\H(\b{r}_1,\b{r}_2) = w_\ee(\b{r}_1,\b{r}_2)$ is the Hartree kernel, $f_\H^{\lr,\HF}(\b{r}_1,\b{r}_1';\b{r}_2,\b{r}_2') = -(1/2)\delta(\b{r}_1-\b{r}_2) \delta(\b{r}_1'-\b{r}_2') w_\ee^{\lr}(\b{r}_1,\b{r}_1')$ is the nonlocal HF exchange kernel, and $f_\xc^{\sr}[\rho_{\gamma_0}](\b{r}_1,\b{r}_2)$ is the short-range exchange-correlation kernel, which for the LDA [Eq.~(\ref{vxcsr})] takes the local form
\begin{eqnarray}
f_\xc^{\sr}[\rho_{\gamma_0}](\b{r}_1,\b{r}_2) = \delta(\b{r}_1-\b{r}_2)  \left. \frac{\partial^2 \bar{e}_{\xc,\text{UEG}}^\sr(\rho,\mu)}{\partial \rho^2} \right|_{\rho=\rho_{\gamma_0}(\b{r}_1)}.
\label{fxcsr}
\end{eqnarray}

From the form of the perturbation in Eq.~(\ref{vextt}), we can write $\psi_i^{(1)}$ as
\begin{eqnarray}
\psi_i^{(1)}(\b{r},t) = \left[ \psi_i^{(+)}(\b{r},\omega)  e^{-\i \omega  t} + \psi_i^{(-)}(\b{r},\omega)    e^{+\i \omega t} \right] e^{\eta t},
\label{psii1rt}
\end{eqnarray}
which, after insertion into Eq.~(\ref{eqforpsii1}), gives the TDRSH or TDLRSH Sternheimer equations for $\psi_i^{(+)}$ and $\psi_i^{(-)}$, written in a common form,
\begin{widetext}
\begin{eqnarray}
\!\! \left(\pm \omega +\i \eta +\varepsilon_i \right) \psi_i^{(\pm)}(\b{r}_1,\omega) &=& \! \! \int_{\mathbb{R}^3} \! h[\gamma_0](\b{r}_1,\b{r}_1') \psi_i^{(\pm)}(\b{r}_1',\omega) \d\b{r}_1'
\nonumber\\
&& + \!\! \int_{\mathbb{R}^9} f_\Hxc[\gamma_0](\b{r}_1,\b{r}_1';\b{r}_2,\b{r}_2') \gamma^{(\pm)}(\b{r}_2,\b{r}_2',\omega) \varphi_i(\b{r}_1') \d\b{r}_1' \d\b{r}_2 \d\b{r}_2'
+v_\text{ext}^{(1)}(\b{r}_1) \varphi_i(\b{r}_1), \;\;
\label{Sternheimerrealspace}
\end{eqnarray}
\end{widetext}
where we have introduced $v_\text{ext}^{(1)}(\b{r}) = \b{r} \cdot \b{e}$ and
\begin{eqnarray}
\gamma^{(\pm)}(\b{r},\b{r}',\omega) = 2\sum_{i=1}^{N/2} \left[\psi_i^{(\pm)}(\b{r},\omega) \varphi_i^*(\b{r}') + \varphi_i(\b{r}) \psi_i^{(\mp)*}(\b{r}',\omega) \right].
\nonumber\\
\label{gammapm}
\end{eqnarray}
As long as $\eta >0$, the solutions $\psi_i^{(\pm)}$ of Eq.~(\ref{Sternheimerrealspace}) are properly square-integrable for any fixed frequency $\omega$. Note that if we had introduced $\psi_i^{(-)*}(\b{r},\omega)$ in place of $\psi_i^{(-)}(\b{r},\omega)$ in Eq.~(\ref{psii1rt}), like in Ref.~\onlinecite{HofSchKum-JCP-18}, then we would have obtained an equation similar to Eq.~(\ref{Sternheimerrealspace}) but with imaginary shifts $\pm\i \eta$.

The photoexcitation/photoionization cross section can then be calculated as~\cite{HofSchKum-JCP-18}
\begin{eqnarray}
\sigma(\omega) = \lim_{\eta \to 0^+} \frac{4\pi \omega}{c} \text{Im}[\alpha(\omega + \i \eta)],
\label{sigma}
\end{eqnarray}
where $c = 137.036$ a.u. is the speed of light and $\alpha(\omega)$ is the spherically averaged dipole polarizability given by
\begin{eqnarray}
\alpha(\omega + \i \eta) = - \frac{1}{3} \sum_{a\in \{x,y,z\}} \int_{\mathbb{R}^3} (\b{r} \cdot \b{u}_a) \; \rho^{(+)} (\b{r},\omega) \d\b{r},
\label{alpha}
\end{eqnarray}
where $\b{u}_a$ is the unit vector along the direction $a$ and $\rho^{(+)}(\b{r},\omega) = \gamma^{(+)} (\b{r},\b{r},\omega)$ is the Fourier component of the first-order density at frequency $\omega + \i \eta$ of the first-order change of the density [Eq.~(\ref{gammapm})]. 

Note that, most often, in the derivation of the Sternheimer equations~\cite{SteDecLis-JPB-95,AndBotMarRub-JCP-07,StrLehRubMarLou-INC-12,HofSchKum-JCP-18}, $\psi_i^{(\pm)}$ are in fact defined so that they are orthogonal to $\varphi_i$, which lead to adding the projector operator onto the space orthogonal to $\varphi_i$ acting on the last two terms in Eq.~(\ref{Sternheimerrealspace}). While this is probably a good choice for numerical calculations, it is not mandatory for the theoretical derivation, as discussed in Ref.~\onlinecite{HofSchKum-JCP-18}, and in any case it leads to the same observable quantities such as the polarizability in Eq.~(\ref{alpha}).

\subsection{Boundary conditions on a finite domain}

It is instructive to consider the behavior of Eq.
(\ref{Sternheimerrealspace}) far away from the atom or molecule. In
this case, the equation reduces to
\begin{align}
  \label{eq:model_equation}
  (\varepsilon_{i} \pm \omega + i\eta - \frac{1}{2} \bm{\nabla}^{2}) \psi_{i}^{(\pm)}(\b r,\omega) \approx 0.
\end{align}
In the limit $\eta \to 0^{+}$, this equation has very different
behavior depending on the sign of $\varepsilon_{i} \pm \omega$.
When $\varepsilon_{i} \pm \omega  < 0$ (below
the ionization threshold), the solutions decay exponentially fast at infinity. By contrast, when
$\varepsilon_{i} \pm \omega > 0$ (above the ionization threshold), they are
oscillatory (behaving like free outgoing waves). As shown in
the Appendix for the case of atoms, this analysis can be refined to take into account the presence of
the effective long-range Coulomb potential and find the exact asymptotic behavior of $\psi_{i}^{(\pm)}$.

Let us then consider the case where the numerical computation is truncated to a bounded spatial domain $\Omega \subset \mathbb{R}^3$.
Using zero (Dirichlet) boundary conditions for $\psi_{i}^{(\pm)}$ on the
boundary $\partial \Omega$ is clearly not appropriate when
$\varepsilon_{i} \pm \omega  > 0$, and indeed will artificially discretize the
electronic continuum, in turn discretizing the photoionization
spectrum. Instead, following the general philosophy of the
Dirichlet-to-Neumann approach (see, e.g., Refs.~\onlinecite{AntLorTan-MP-17,SzmBie-PRA-04}), we can use frequency-dependent boundary
conditions that gives accurate results even for a relatively small domain and $\eta=0$.
For this, we just need to find analytically an approximation of $\psi_{i}^{(\pm)}$ which is valid outside of $\Omega$, and then
require the normal derivative of $\psi_{i}^{(\pm)}$ to match on $\partial \Omega$ between the interior
and exterior, which yields a nonlocal Robin boundary condition of the form
\begin{align}
  \label{eq:boundary_condition}
\forall \b r \in \partial\Omega,\;    \b{n}(\b{r}) \!\cdot \!\bm{\nabla} \psi_{i}^{(\pm)}(\b{r},\omega) = \int_{\partial \Omega} K_i(\b r, \b r'; \pm \omega) \psi_{i}^{(\pm)}(\b r', \omega) \d \b r',
\end{align}
where $\b{n}(\b{r})$ is the outward normal vector to the surface $\partial\Omega$ at point $\b{r}$. In the case of atoms, the Dirichlet-to-Neumann kernel $K_i(\b r, \b r'; \pm \omega)$ reduces to a simple local radial form (see Appendix).

\vspace{0.5cm}
\subsection{Linear-response Sternheimer equations in a basis set}
\label{sec:sternheimer_basis_set}

Let us introduce now a finite (non-orthogonal) basis set
$\{\chi_\nu\}_{\nu=1,...,M} \subset H^1(\Omega,\mathbb{C})$ (where $H^1$ is the first-order Sobolev space) made of
$M$ basis functions (whose behavior on $\partial\Omega$ is arbitrary) to expand the occupied orbitals
\begin{eqnarray}
\varphi_j(\b{r}) = \sum_{\nu=1}^{M} c_{j\nu} \chi_\nu(\b{r}),
\label{phijexpand}
\end{eqnarray}
and their first-order changes
\begin{eqnarray}
\psi_j^{(\pm)}(\b{r},\omega) = \sum_{\nu=1}^{M} c_{j\nu}^{(\pm)}(\omega) \chi_\nu(\b{r}),
\label{psijexpand}
\end{eqnarray}
where $c_{j\nu}$ and $c_{j\nu}^{(\pm)}(\omega)$ are (generally complex-valued) coefficients labeled with the composite index $j\nu \equiv (j,\nu) \in  \llbracket  1, N/2\rrbracket \times \llbracket  1, M \rrbracket$. Integrating Eq.~(\ref{Sternheimerrealspace}) against a basis function $\chi_\mu^*$, and using the expansions of Eqs.~(\ref{phijexpand}) and~(\ref{psijexpand}), 
leads directly to the basis-set Sternheimer equations in the following block matrix form
\begin{eqnarray}
\left( \begin{array}{cc}
\bm{\Lambda}(\omega) & \b{B} \\
\b{B}^* & \b{\Lambda}(-\omega)^*
\end{array} \right)
\left( \begin{array}{c}
\b{c}^{(+)}(\omega)  \\
\;\b{c}^{(-)}(\omega)^* \\
\end{array} \right)
=-\left( \begin{array}{c}
\b{V}  \\
\;\b{V}^* \\
\end{array} \right),
\label{Sternheimermatrix}
\end{eqnarray}
which must be solved at each given frequency $\omega$ for $\b{c}^{(+)}(\omega)$ and $\b{c}^{(-)}(\omega)^*$ which are the column vectors of components $c^{(+)}_{j\nu}(\omega)$ and $c^{(-)}_{j\nu}(\omega)^*$, respectively. In Eq.~(\ref{Sternheimermatrix}), $\b{V}$ is the column vector of components $V_{i\mu}= \b{e} \cdot \sum_{\nu=1}^M \b{d}_{\mu,\nu} c_{i\nu}$ where $\b{d}_{\mu,\nu}= \int_{\Omega} \chi_\mu^*(\b{r}) \b{r} \chi_\nu(\b{r}) \d\b{r}$ are the dipole-moment integrals, and $\bm{\Lambda}(\pm\omega)$ and $\b{B}$ are square matrices with elements
\begin{eqnarray}
\Lambda_{i\mu,j\nu}(\pm\omega) = \delta_{i,j} \left( h_{i,\mu,\nu}(\pm \omega) - (\varepsilon_{i} \pm \omega +\i \eta) S_{\mu,\nu} \right) 
\nonumber\\
+2 \sum_{\l=1}^{M} \sum_{\sigma=1}^{M} 
\; c_{i\sigma} c_{j\l}^* F_{\mu,\l,\sigma,\nu},
\label{Lambda}
\end{eqnarray}
and
\begin{eqnarray}
B_{i\mu,j\nu} &=& 2 \sum_{\l=1}^{M} \sum_{\sigma=1}^{M} c_{i\sigma} c_{j\l} F_{\mu,\nu,\sigma,\l}.
\label{B}
\end{eqnarray}
In Eq.~(\ref{Lambda}), $S_{\mu,\nu} = \int_{\Omega} \chi_\mu^*(\b{r}) \chi_\nu(\b{r}) \d\b{r}$ are the overlap integrals over the basis functions, and $h_{i,\mu,\nu}(\pm \omega)$ are the matrix elements of the RSH or LRSH Hamiltonian
\begin{eqnarray}
h_{i,\mu,\nu}(\pm \omega)       &=& t_{i,\mu,\nu}(\pm \omega) + v_{\mu,\nu}
\\
 &&+ \sum_{\l=1}^{M} \sum_{\sigma=1}^{M} P_{\sigma,\l} \left( w_{\mu,\l,\nu,\sigma} - \frac{1}{2} w_{\mu,\l,\sigma,\nu}^\lr \right) + v_{\mu,\nu}^\sr,\nonumber
\label{hmunu}
\end{eqnarray}
where $t_{i,\mu,\nu}(\pm \omega)$ are the kinetic integrals including the boundary condition [see Eq.~(\ref{eq:kinetic_integrals}) below], $v_{\mu,\nu} = \int_{\Omega} \chi_\mu^*(\b{r}) v_\text{ne}(\b{r}) \chi_\nu(\b{r}) \d\b{r}$ are the nuclei-electron integrals, $P_{\sigma,\l} = 2 \sum_{i=1}^{N/2} c_{i\sigma} c_{i\l}^*$ are the elements of the density matrix, $w_{\mu,\l,\nu,\sigma} = \int_{\Omega^2} \chi_\mu^*(\b{r}_1) \chi_\l^*(\b{r}_2) w_\ee(\b{r}_1,\b{r}_2) \chi_\nu(\b{r}_1) \chi_\sigma(\b{r}_2) \d\b{r}_1 \d\b{r}_2$ and $w_{\mu,\l,\sigma,\nu}^\lr = \int_{\Omega^2} \chi_\mu^*(\b{r}_1) \chi_\l^*(\b{r}_2) w_\ee^\lr(\b{r}_1,\b{r}_2) \chi_\sigma(\b{r}_1) \chi_\nu(\b{r}_2) \d\b{r}_1 \d\b{r}_2$ are the Coulombic and long-range two-electron integrals, respectively, and $v_{\mu,\nu}^\sr  = \int_{\Omega} \chi_\mu^*(\b{r}) v_\xc^\sr(\b{r}) \chi_\nu(\b{r}) \d\b{r}$ are the short-range exchange-correlation potential integrals. In Eqs.~(\ref{Lambda}) and~(\ref{B}), $F_{\mu,\l,\sigma,\nu}$ are the matrix elements of the Hartree-exchange-correlation kernel $f_\Hxc[\gamma_0]$
\begin{widetext}
\begin{eqnarray}
F_{\mu,\l,\sigma,\nu} &=& \int_{\Omega^4} \chi_\mu^*(\b{r}_1) \chi_\l^*(\b{r}_2) \; f_\Hxc[\gamma_0](\b{r}_1,\b{r}_1';\b{r}_2,\b{r}_2') \; \chi_\sigma(\b{r}_1') \chi_\nu(\b{r}_2') \; \d\b{r}_1 \d\b{r}_1' \d\b{r}_2 \d\b{r}_2'
\nonumber\\
&=& w_{\mu,\l,\sigma,\nu} - \frac{1}{2} w_{\mu,\l,\nu,\sigma}^\lr + f^\sr_{\mu,\l,\sigma,\nu},
\end{eqnarray}
where $f^\sr_{\mu,\l,\sigma,\nu} = \int_{\Omega^2} \chi_\mu^*(\b{r}_1) \chi_\l^*(\b{r}_2) f_\xc^\sr[\rho_{\gamma_0}](\b{r}_1,\b{r}_2) \chi_\sigma(\b{r}_1) \chi_\nu(\b{r}_2) \d\b{r}_1 \d\b{r}_2$ are the short-range exchange-correlation kernel integrals.
\end{widetext}

To obtain the expression of the kinetic integrals, we start from the kinetic contribution in Eq.~(\ref{Sternheimerrealspace}), projected on the basis function $\chi_\mu^*$, and perform an integration by parts
\begin{eqnarray}
-\frac{1}{2} \int_{\Omega} \chi_\mu^*(\b{r}) \bm{\nabla}^2 \psi^{(\pm)}_i(\b{r},\omega) \d\b{r}
= \frac{1}{2} \int_{\Omega} \bm{\nabla} \chi_\mu^*(\b{r}) \cdot \bm{\nabla} \psi^{(\pm)}_i(\b{r},\omega) \d\b{r}
\nonumber\\
-\frac{1}{2} \int_{\partial\Omega} \chi_\mu^*(\b{r}) \; \b{n}(\b{r}) \!\cdot \!\bm{\nabla} \psi_{i}^{(\pm)}(\b{r},\omega) \d\b{r}.
\nonumber\\
\end{eqnarray}
Using the boundary condition in Eq.~(\ref{eq:boundary_condition}), the surface term can then be expressed as
\begin{eqnarray}
-\frac{1}{2} \int_{\partial\Omega} \chi_\mu^*(\b{r})\; \b{n}(\b{r}) \!\cdot \!\bm{\nabla} \psi_{i}^{(\pm)}(\b{r},\omega) \d\b{r} = \phantom{xxxxxxxxx}
\nonumber\\
-\frac{1}{2} \int_{\partial\Omega^2} \chi_\mu^*(\b{r}) K_i(\b r, \b r'; \pm \omega) \psi_{i}^{(\pm)}(\b r', \omega) \d\b{r} \d \b r'.
\end{eqnarray}
After expanding $\psi_{i}^{(\pm)}$ in the basis set, we thus obtain the expression of the kinetic integrals
\begin{eqnarray}
  \label{eq:kinetic_integrals}
  t_{i,\mu,\nu}(\pm \omega) &=& \frac 1 2 \int_{\Omega} \nabla \chi_\mu^*(\b{r})\cdot \nabla \chi_\nu(\b{r}) \d\b{r} \\
  &&- \frac{1}{2} \int_{\partial\Omega^{2}} \chi_{\mu}^*(\b r) K_i(\b r,\b r';\pm \omega) \chi_{\nu}(\b r')\d\b{r} \d\b{r'} .\nonumber
\end{eqnarray}
In the limit of a complete basis set, this correctly imposes Eq.~(\ref{eq:boundary_condition}) on $\psi_{i}^{(\pm)}$.

To see this, it is instructive to consider the simplified problem of solving
the one-dimensional differential equation $-\psi''(x) + V(x) \psi(x) = F(x)$ on
the interval $(-1,1)$, for some real-valued potential $V$ and source term $F$,
and boundary conditions $\psi'(\pm 1) = K_{\pm} \; \psi(\pm 1)$. Solving this problem is equivalent to requiring
that $\psi$ satisfies the variational equation
$\int_{-1}^1 \chi'(x) \psi'(x) \d x - [\chi(1) K_{+} \psi(1) - \chi(-1) K_{-} \psi(-1)] + \int_{-1}^1 \chi(x) V(x) \psi(x) \d x = \int_{-1}^1 \chi(x) F(x) \d x$
for all $\chi \in H^{1}(-1,1)$.
Choosing for $\chi$ a function that is zero on $(-1,1-1/n)$ and
ramps up linearly to $1$ at $x\to 1$, we obtain by passing to the limit
$n\to \infty$ in the variational equation that $\psi'(1) = K_{+} \psi(1)$.
Similarly, we can choose the function $\chi$ so that $\psi'(-1)=K_{-}\psi(-1)$.

The Sternheimer matrix equations in Eq.~(\ref{Sternheimermatrix}) have essentially the same form as the well-known TDDFT Casida equations~\cite{Cas-INC-95,Cas-JMS-09} except that the latter are normally written in the orthogonal basis of the occupied and virtual orbitals whereas the present Sternheimer matrix equations are written in an arbitrary non-orthogonal basis set.

Finally, in the basis set, the dipole polarizability takes the form
\begin{eqnarray}
\alpha(\omega + \i \eta) &=& - \frac{1}{3} \sum_{a\in \{x,y,z\}} \sum_{\mu=1}^M \sum_{\nu=1}^M \left( P_{\mu,\nu}^{(+)}(\omega) \b{d}_{\nu,\mu} + P_{\mu,\nu}^{(-)}(\omega)^* \b{d}_{\nu,\mu}^* \right) \cdot \b{u}_a,
\nonumber\\
\end{eqnarray}
where $P_{\mu,\nu}^{(\pm)}(\omega)=2\sum_{i=1}^{N/2} c_{i\mu}^{(\pm)}(\omega) c_{i\nu}^*$.

\subsection{B-spline basis set for atoms}

We specialize now in the case of atoms with radial ground-state densities. Spherical symmetry permits to write the occupied orbitals as
\begin{eqnarray}
\varphi_{i}(\b{r}) = \frac{R_i(r)}{r} Y_{\ell_i}^{m_i}(\theta,\phi),
\end{eqnarray}
with radial functions $R_i$ and spherical harmonics $Y_{\ell_i}^{m_i}$. For a dipole interaction with a $z$-polarized electric field, i.e. $v_\text{ext}^{(1)}= \b{r} \cdot \b{u}_z$, 
the corresponding first-order orbital changes are of the form
\begin{eqnarray}
\psi_{i}^{(\pm)}(\b{r},\omega) = \sum_{\ell\in {\cal L}_{i}} \frac{R_{i,\ell}^{(\pm)}(r,\omega)}{r} Y_{\ell}^{m_i}(\theta,\phi),
\label{psijrmratoms}
\end{eqnarray}
with ${\cal L}_i  = \{ \ell_i -1, \ell_i +1 \}$ for $\ell_i\ge 1$ and ${\cal L}_i  = \{ \ell_i +1 \}$ for $\ell_i=0$.

We expand the radial functions in a basis set of $M_\s$ B-spline functions~\cite{Boo-BOOK-78,BacCorDecHanMar-RPP-01} $\{B_\nu\}_{\nu=1,...,M_\s}$ of order $k_\s$
\begin{eqnarray}
 R_{i}(r)=\sum_{\nu=1}^{M_\s}c_{i\nu} B_\nu(r),
\end{eqnarray}
\begin{eqnarray}
 R_{i,\ell}^{(\pm)}(r,\omega)=\sum_{\nu=1}^{M_\s}c_{i\nu,\ell}^{(\pm)}(\omega) B_\nu(r).
\label{Ripmexpand}
\end{eqnarray}
To completely define a basis of B-spline functions, a non-decreasing sequence of $M_\s+k_\s$ knot points $\{r_p\}_{p=1,...,M_\s+k_\s}$ (some knot points are possibly coincident) must be given. The B-spline function $B_\nu(r)$ is non zero only on the supporting interval $[r_\nu,r_{\nu+k_\s}]$ (containing $k_\s+1$ consecutive knot points) and is a piecewise function composed of polynomials of degree $k_\s-1$ (one polynomial in between two consecutive non-coincident knot points) with continuous first $k_\s-m-1$ derivatives across each knot of multiplicity $m$. We have followed the standard choice of taking the first and the last knots to be $k_\s$-fold degenerate, i.e. $r_1 = r_2 = \cdots = r_{k_\s} = r_{\text{min}}=0$ and $r_{{M_\s+1}} = r_{{M_\s+2}} = \cdots = r_{{M_\s+k_\s}}= r_{\text{max}}$, while the multiplicity of the other knots is unity. We thus need $k_\s \ge 3$ in order to have basis functions with at least $C^1$ regularity. The spatial grid spacing was chosen to be constant in the whole radial space between two consecutive non-coincident points and is given by $\Delta r = r_{\text{max}}/(M_\s-k_\s+1)$.

At $r=0$, the appropriate boundary conditions are $R_i(r=0)=0$ and
$R_{i,\ell}^{(\pm)}(r=0,\omega)=0$, which can be easily imposed by removing the
first B-spline function which is the only one non vanishing at $r=0$.
At $r=r_\text{max}$, for the occupied orbitals, which decay exponentially 
fast at infinity, we can also use zero
boundary conditions, i.e. $R_i(r=r_\text{max})=0$, which can be imposed
by just removing the last B-spline function in the ground-state
calculation. For the first-order orbital changes
$R_{i,\ell}^{(\pm)}$, as shown in the Appendix, the radial symmetry
simplifies the nonlocal boundary condition
in Eq.~(\ref{eq:boundary_condition}) to the local Robin boundary condition
\begin{align*}
\left.  \frac{\d R_{i,\ell}^{(\pm)}(r,\omega)}{\d r} \right|_{r=r_{\text{max}}} = b_{i,\ell}(\pm\omega) R_{i,\ell}^{(\pm)}(r_{\text{max}}),
\end{align*}
and the kinetic integrals [Eq.~(\ref{eq:kinetic_integrals})] for $R_{i,\ell}^{(\pm)}$ become
\begin{eqnarray}
t_{i,\ell,\mu,\nu}(\pm\omega) &=& \frac{1}{2}  \int_0^{r_\text{max}} \frac{\d B_{\mu}(r)}{\d r} \frac{\d B_{\nu}(r)}{\d r} \d r  
\nonumber\\
&&-\delta_{\mu,M_\s} \delta_{\nu,M_\s}\frac{b_{i,\ell}(\pm\omega)}{2} B_{M_\s}(r_\text{max})^2,
\label{timunu}
\end{eqnarray}
where we have used the fact that only the last B-spline basis function is non zero at $r_\text{max}$. 
The complex-valued function $b_{i,\ell}$ is given in Eqs.~(\ref{blargeomega}),~(\ref{bsmallomega}), and~(\ref{bmomega}).
In this way, the boundary condition is imposed without modification of
the basis set. This method is somewhat simpler than the procedure to
impose boundary conditions described in
Refs.~\onlinecite{MahSub-BOOK-90,SteDecLis-JPB-95}, but conceptually
similar to the procedure described in
Refs.~\onlinecite{NakYab-JCP-01,YabNakIwaBer-PSS-06} for grid-based
TDDFT and in
Refs.~\onlinecite{Zat-CPC-06,BurNobBur-INC-07,ZatBar-JPB-13} in the
context of the R-matrix method.

\subsection{Further computational details}

We apply the present theory to the Be atom ($N=4$). We use $M_\s=50$ B-spline basis functions of order $k_\s=8$ and a maximal radius of $r_{\text{max}} = 25$ bohr. The occupied orbitals are of symmetry s ($\ell_i=0$, $m_i=0$) and the perturbed orbitals are of symmetry p$_z$ ($\ell=1$). 

Radial integrals over B-spline functions are calculated using a Gauss-Legendre quadrature~\cite{BacCorDecHanMar-RPP-01}. We use the integration-cell algorithm~\cite{QiuFro-JCP-99} to calculate the Coulomb two-electron integrals, and an extension of it~\cite{ZapLupTou-JCP-19} for the long-range two-electron integrals. For the case of LRSH, having a position-dependent range-separation parameter in the long-range two-electron integrals does not introduce any complications since the integration-cell algorithm uses a Gaussian quadrature. Thanks to the locality of the B-spline basis functions, the construction of the matrices $\bm{\Lambda}(\omega)$ and $\b{B}$ in Eq.~(\ref{Sternheimermatrix}) scales as $O(N^2 M_\s^2)$, instead of the straightforward scaling $O(N^2 M^4)$ that would be obtained for an arbitrary basis set. The Sternheimer equations are usually solved iteratively~\cite{SteDecLis-JPB-95,AndBotMarRub-JCP-07,HofSchKum-JCP-18}. However, we found that the simple iterative scheme of Ref.~\onlinecite{SteDecLis-JPB-95} starting from the bare response, i.e. $f_\Hxc[\gamma_0]=0$, works well for TDLDA but often does not converge for TDHF because the bare HF response is too bad an approximation to the TDHF response. Although this could be cured with more sophisticated iterative methods, on this simple system we found it expedient to simply solve Eq.~(\ref{Sternheimermatrix}) using a standard dense LU-decomposition linear solver.

Having a non-zero imaginary-shift parameter $\eta$ is necessary to obtain a non-zero absorption cross section for bound states~\cite{HofSchKum-JCP-18}. However, it is not necessary for the continuum part of the spectrum. Indeed, the imaginary part of the dipole polarizability involved in the photoionization cross section in Eq.~(\ref{sigma}) is not zero in any case due to the complex-valued function $b$ introduced when imposing the boundary condition in Eq.~(\ref{timunu}). Moreover, since we are interested in obtaining the precise lineshape of the core resonances, we use $\eta=0$ to avoid artificially broadening these resonances. 

As shown in the Appendix, the use of the accurate boundary conditions in Eq.~(\ref{timunu}) is essential to obtain photoionization cross section without spurious oscillations even with the relatively small $r_{\text{max}}$ that we use.

\section{Results and discussion}
\label{sec:results}

We now show and discuss the results on the Be atom.

\subsection{Orbital energies}
\label{sec:orbenergies}

\begin{figure}
\includegraphics[scale=0.35,angle=-90]{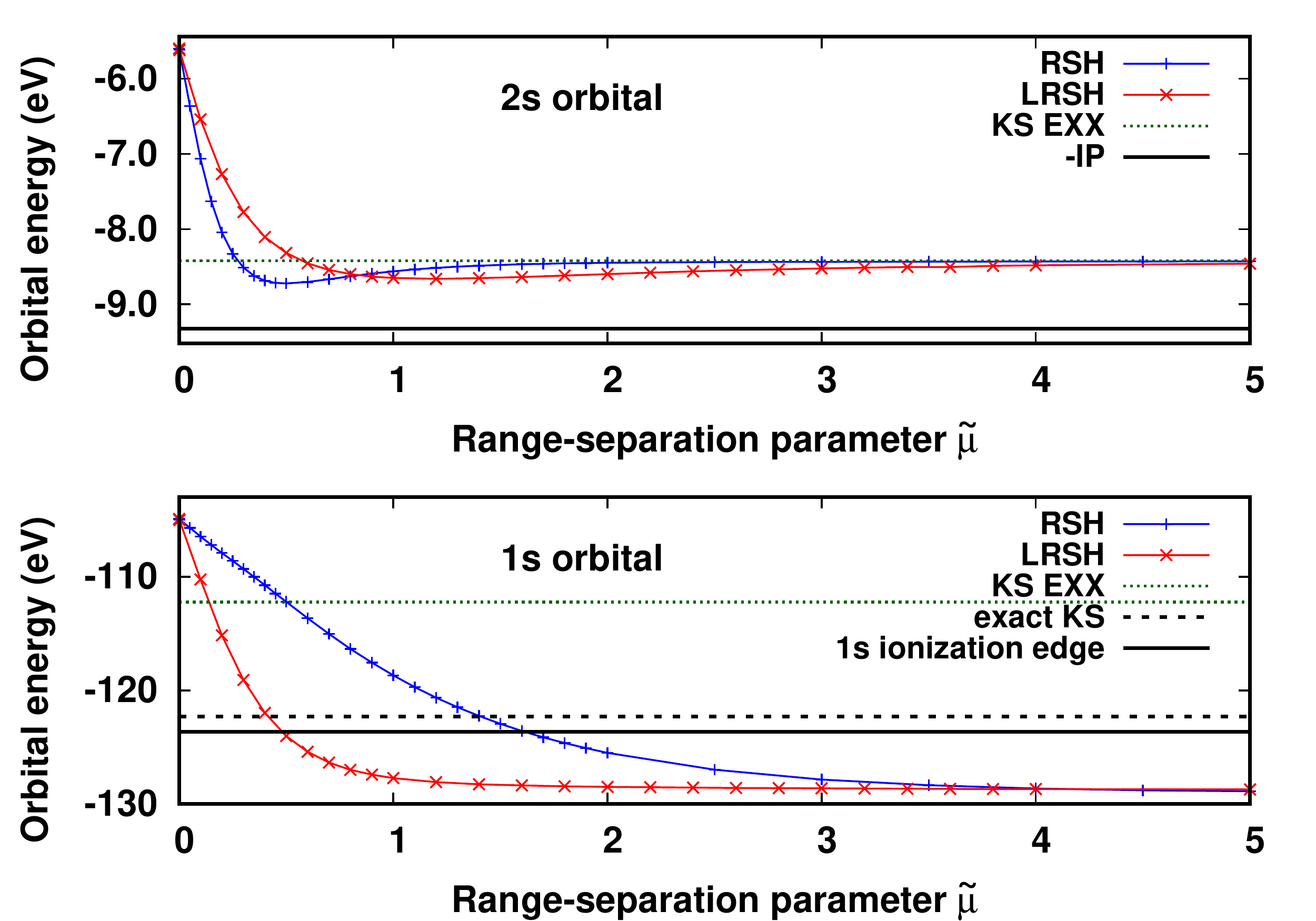}
\caption{RSH and LRSH 1s and 2s orbital energies of the Be atom as a function of the adimensional range-separation parameter $\tilde{\mu}$. As references, the opposite of the experimental IP (9.323 eV) and of the 1s ionization edge (123.64 eV)~\cite{NIST-INC-21} are indicated, as well as the KS EXX 1s and 2s orbital energies (112.24 eV and 8.422 eV)~\cite{GluFesPol-TCA-09} and the exact KS 1s orbital energy (122.29 eV)~\cite{ChoGriBae-JCP-02}.}
\label{fig:orbitalenergies}
\end{figure}

Figure~\ref{fig:orbitalenergies} shows the RSH and LRSH 1s and 2s orbital energies as a function of the adimensional range-separation parameter $\tilde{\mu}$. Also indicated are the opposite of the experimental ionization potential (IP) (9.323 eV) and of the 1s ionization edge (123.64 eV)~\cite{NIST-INC-21}, as well as the KS exact exchange (EXX) 1s and 2s orbital energies (112.24 eV and 8.422 eV)~\cite{GluFesPol-TCA-09} and the exact KS 1s orbital energy (122.29 eV)~\cite{ChoGriBae-JCP-02}. 

For $\tilde{\mu}=0$, both RSH and LRSH reduce to the standard KS scheme. If we were to use the exact exchange-correlation potential, the exact KS 2s orbital energy would of course be exactly the opposite of the experimental IP. The exact KS 1s orbital energy is not equal to the opposite of the experimental 1s ionization edge but can be considered as an approximation to it~\cite{ChoGriBae-JCP-02}. Here, due to the use of the LDA for the exchange-correlation potential, the 1s and 2s orbital energies at $\tilde{\mu}=0$ are too high by 17.4 and 3.7 eV, respectively. Too high orbital energies are often attributed to the self-interaction error of the LDA. We see, however, that even KS EXX (which is without self-interaction error) gives 1s and 2s orbital energies that are too high by 10.1 and 0.9 eV, respectively, with respect to exact KS. The latter errors are of course due to the missing KS correlation potential. For $\tilde{\mu} \to \infty$, both RSH and LRSH reduce to standard HF. In comparison with the experimental values, the HF 2s orbital energy is too high by 0.9 eV and the HF 1s orbital energy is too low by 5.2 eV. In the context of Green-function theory, the latter errors are due to the missing correlation self-energy contribution. Interestingly, we see that KS EXX and HF give nearly identical 2s orbital energies. Recalling the fact that KS EXX and HF are identical for two electrons in a single orbital, this likely means that the two valence electrons can be considered as nearly independent from the core electrons for calculating the 2s orbital energy. By contrast, the KS EXX and HF give quite different 1s orbital energies, differing by as much as 16.5 eV. This must mean that the valence electrons cannot be neglected in the mean-field potential for calculating the 1s orbital energy.

For $\tilde{\mu} \neq 0$, even if we were to use the exact short-range exchange-correlation potential, the RSH and LRSH 2s orbital energies would not be exactly equal to the opposite of the experimental IP, since long-range correlation effects are missing in the RSH and LRSH schemes. Also, the RSH and LRSH 1s orbital energies with the exact short-range exchange-correlation potential should not be expected to be exactly equal to the opposite of the experimental 1s ionization edge. However, in the spirit of the optimally tuned range-separated hybrids~\cite{LivBae-PCCP-07,SteKroBae-JACS-09,SteKroBae-JCP-09}, it is reasonable to choose the range-separation parameter $\tilde{\mu}$ in the RSH and LRSH schemes so that the orbital energies are as close as possible to these experimental values. Since for $\tilde{\mu} \gtrsim 0.4$ the RSH and LRSH 2s orbital energies are not very sensitive to the value of $\tilde{\mu}$ and since we are mostly interested in this work in the photoionization spectrum near the 1s ionization edge, we decide to adjust the value of $\tilde{\mu}$ so that the RSH and LRSH 1s orbital energies are equal to opposite of the experimental 1s ionization edge. This gives $\tilde{\mu}=1.608$ for RSH and $\tilde{\mu}=0.478$ for LRSH.

\subsection{Photoionization spectrum}

\begin{figure*}
\includegraphics[scale=0.3,angle=-90]{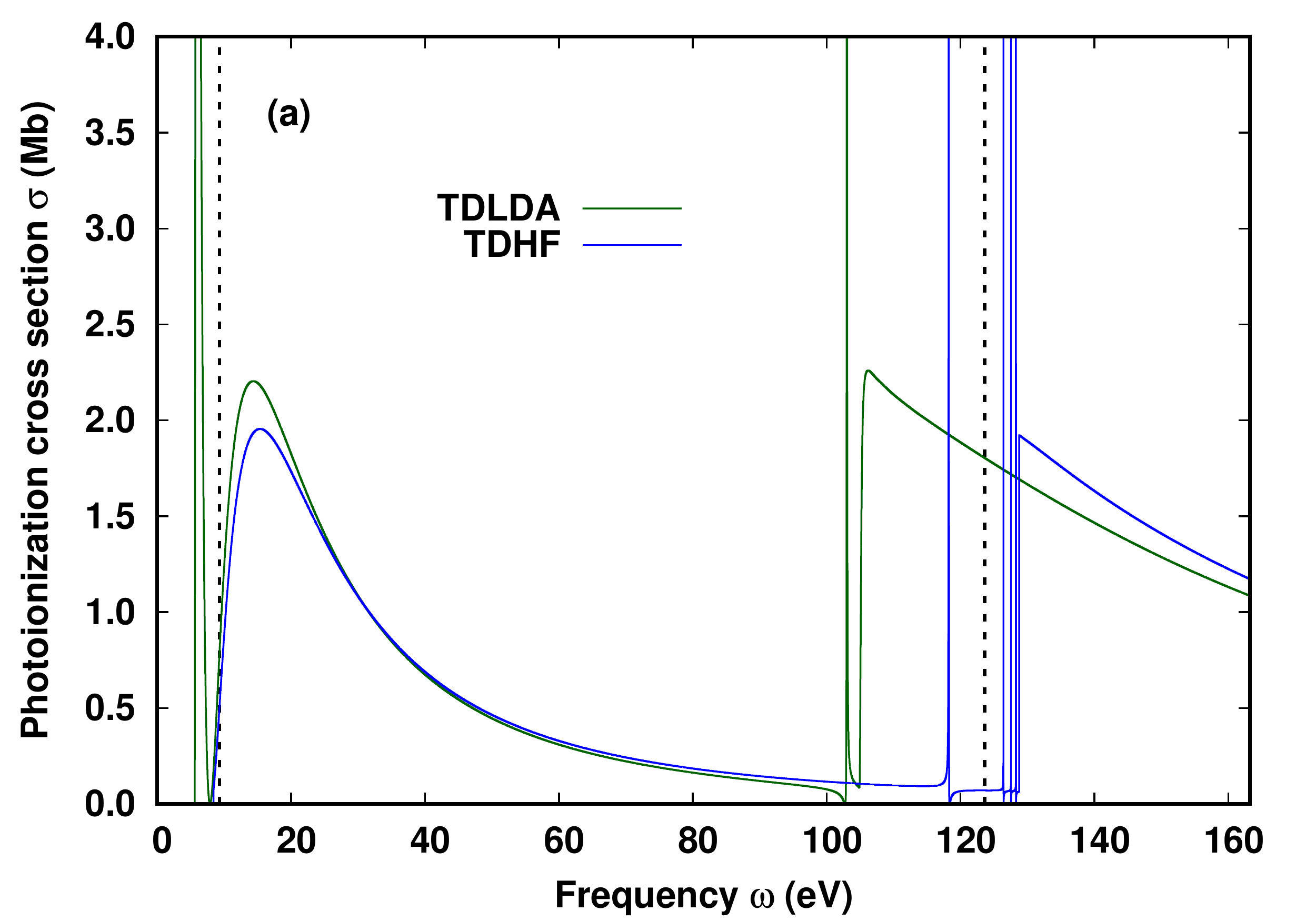}
\includegraphics[scale=0.3,angle=-90]{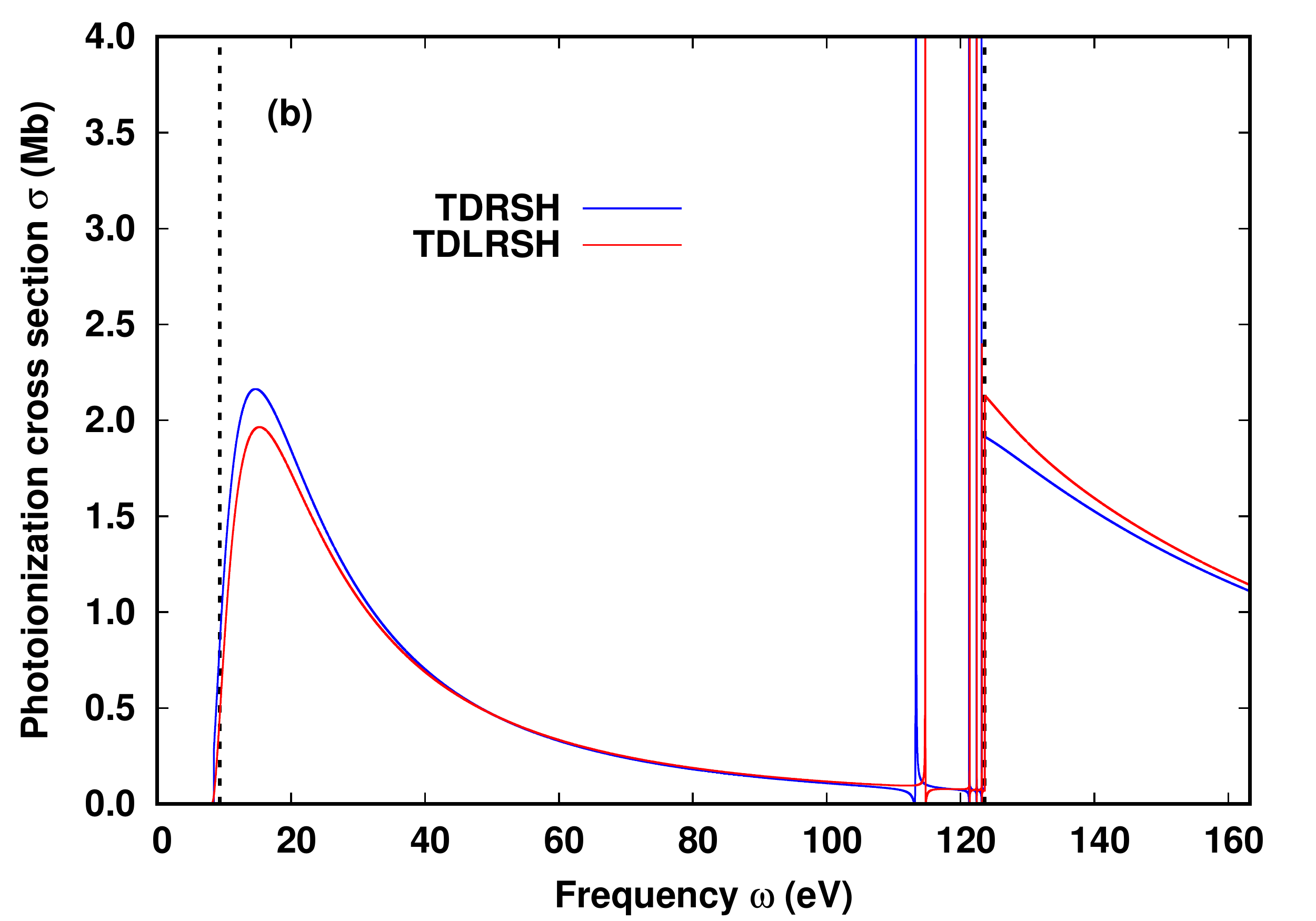}
\caption{Photoionization cross section of the Be atom calculated by (a) TDLDA and TDHF, and by (b) TDRSH and TDLRSH (using the optimal range-separation parameters determined in Section~\ref{sec:orbenergies}, i.e. $\tilde{\mu}=1.608$ for TDRSH and $\tilde{\mu}=0.478$ for TDLRSH). The vertical dashed lines correspond to the experimental IP (9.323 eV) and the 1s ionization edge (123.64 eV)~\cite{NIST-INC-21}.}
\label{fig:photoionization}
\end{figure*}

\begin{figure*}
\includegraphics[scale=0.3,angle=-90]{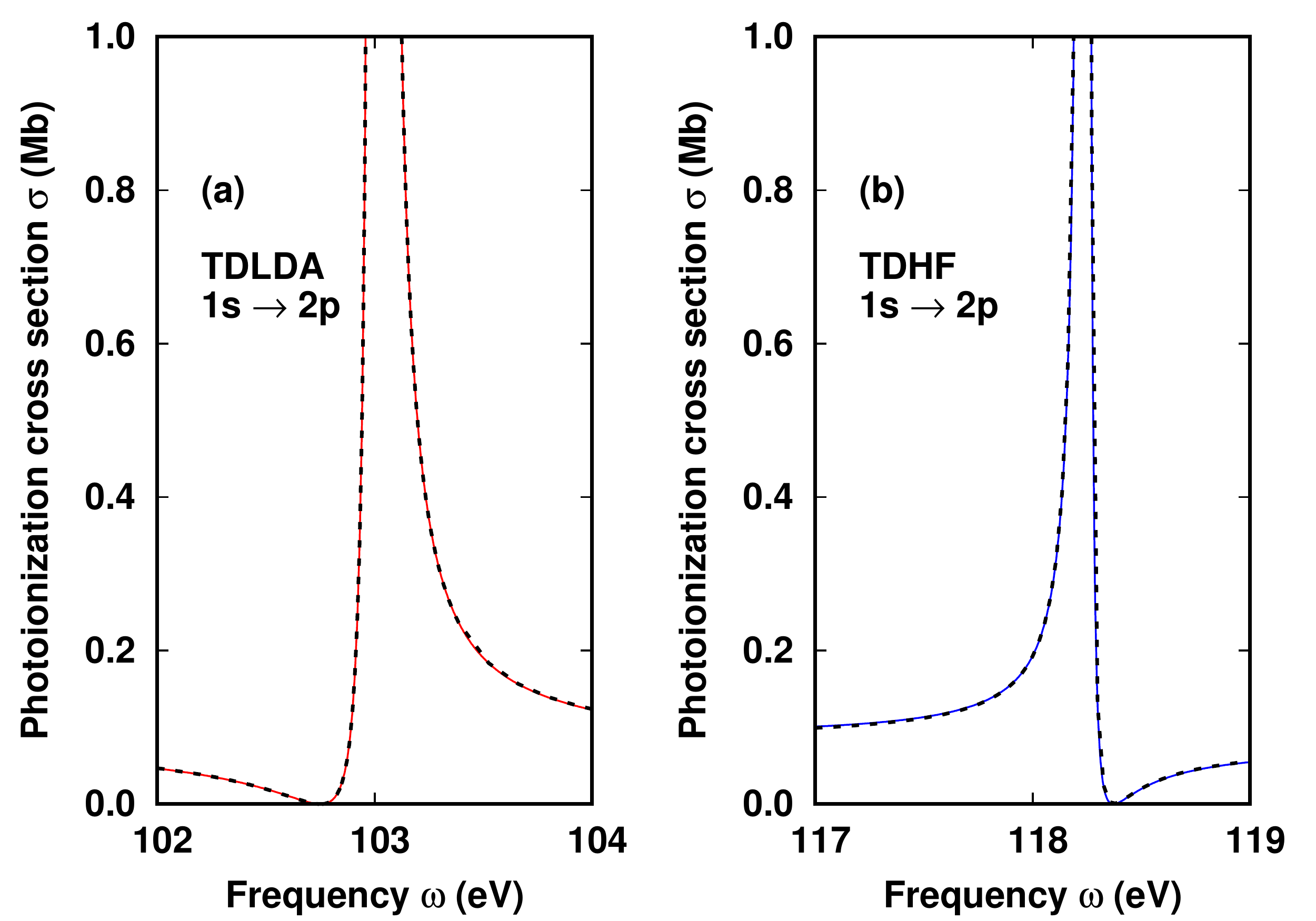}
\includegraphics[scale=0.3,angle=-90]{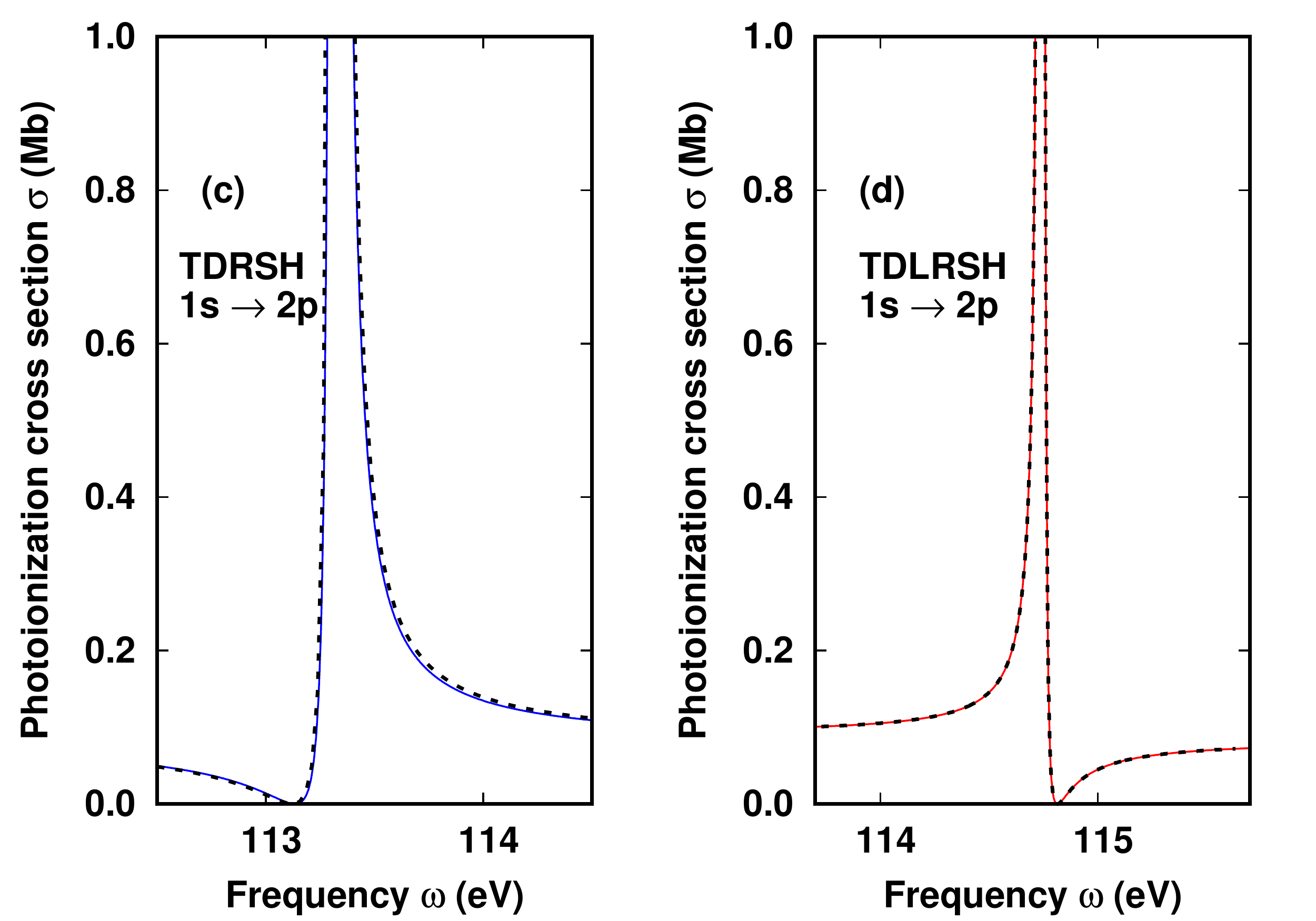}
\caption{Core resonance 1s $\to$ 2p of the Be atom calculated by (a) TDLDA, (b) TDHF, (c) TDRSH, and (d) TDLRSH (using the optimal range-separation parameters determined in Section~\ref{sec:orbenergies}, i.e. $\tilde{\mu}=1.608$ for TDRSH and $\tilde{\mu}=0.478$ for TDLRSH). The dashed lines are fits using Eq.~(\ref{sigmafit}) with the parameters given in Table~\ref{tab:resonance}.}
\label{fig:resonance}
\end{figure*}

Figure~\ref{fig:photoionization} reports the photoionization cross section calculated by TDLDA, TDHF, TDRSH, and TDLRSH (using the optimal range-separation parameters determined in Section~\ref{sec:orbenergies}). We will just comment on the main features of these spectra, without trying to compare with experimental spectra~\cite{KraCal-PRL-87,JanNicTonZheMaz-OC-87,WehLukBlu-PRA-03,WehLukBlu-PRA-05,OlaMenJimWehWhi-PRA-07}. Indeed, the experimental spectra display many more peaks and structures due to double and higher excitations, which are not taken into account in the level of theory that we use.

The TDLDA photoionization spectrum starts at a too low ionization threshold (the same value as the opposite of the LDA 2s orbital energy) and the cross section is zero at the threshold, just like for the He atom~\cite{ZapLupTou-JCP-19}. There is a large peak just above the TDLDA threshold, roughly in the energy range that should correspond to the 2s$\to$3p transition, followed by a Cooper-like minimum~\cite{Coo-PR-62,FanCoo-RMP-68} where the cross section vanishes. We have verified that this minimum is not present in the bare LDA photoionization spectrum and appears when the Hartree kernel is taken into account. The TDLDA 1s ionization edge occurs at a much too low energy, in fact exactly the same value as the opposite of LDA 1s orbital which means that the Hartree-exchange-correlation kernel does not affect this value. The TDLDA photoionization spectrum contains only the first 1s$\to$2p core resonance, the other core single-excited resonances (1s$\to$3p, 1s$\to$4p, etc...) having dissolved into the continuum beyond the 1s ionization edge, due to the exponentially decaying exchange-correlation LDA potential.

The TDHF photoionization spectrum starts at a slightly too low ionization threshold (again the same value as the opposite of the HF orbital 2s energy) and the cross section is very small but not zero (about 0.07 Mb) at the threshold (not shown). The TDHF 1s ionization edge occurs at a too high energy, again the same value as the opposite of HF 1s orbital which means that the HF kernel does not affect this value. In contrast to TDLDA, the TDHF photoionization spectrum not only contains the 1s$\to$2p core resonance but also a series of single-excited core resonances to Rydberg states (1s$\to$3p, 1s$\to$4p, etc...) converging toward the 1s ionization edge. We note that our TDHF photoionization spectrum is in good agreement with previous non-relativistic and relativistic TDHF calculations~\cite{AmuCheZivRad-PRA-76,JohLin-JPB-77,LeeJoh-PRA-80} (note that linear-response TDHF is also called random-phase approximation with exchange).

The TDRSH and TDLRSH photoionization spectra (using the optimal range-separation parameters determined in Section~\ref{sec:orbenergies}) display roughly the same features. They both start at a slightly too low ionization threshold. In both cases, the 1s ionization edge is positioned at the exact value, as expected since it corresponds to the opposite of the 1s orbital energy which has been adjusted at the exact value by tuning the range-separation parameter. Similar to TDHF, both the TDRSH and TDLRSH photoionization spectra display a series of core resonances, the TDLRSH resonances being systematically at higher energies than the TDRSH resonances. In comparison to TDRSH, TDLRSH gives smaller cross sections in the 2s continuum region (near 20 eV) and larger cross sections in the 1s continuum region (above the 1s ionization edge).

\subsection{Core resonances}

\begin{table*}
\label{tab:resonance}
\caption{Resonance energy $E_\text{R}$, resonance width $\Gamma$, Fano asymmetric parameter $q$, total background cross section $\sigma_0$, background ratio parameter $\rho^2$, background linear drift $a$, and maximum value of the cross section at the resonance energy $\sigma(E_\text{R})$ for the 1s$\to$2p and 1s$\to$3p core resonances of the Be atom calculated by TDLDA, TDHF, TDRSH, and TDLRSH (using the optimal range-separation parameters determined in Section~\ref{sec:orbenergies}, i.e. $\tilde{\mu}=1.608$ for TDRSH and $\tilde{\mu}=0.478$ for TDLRSH). In some cases, the background linear drift $a$ was fixed to exactly zero in order to obtain convergence.}
\begin{tabular}{l c c c c c c c}
\hline\hline
                       & $E_\text{R}$ (eV) & $\Gamma$ (meV) & $q$ & $\sigma_0$  (Mb)  & $\rho^2$ & $a$  & $\sigma(E_\text{R})$ (Mb)\\
\hline\\[-0.3cm]
 resonance 1s$\to$2p\\[0.1cm]
  TDLDA                      & 103.0  &  2.347 &  228.3   & 0.081  & 0.998  & -7.73  $\cdot 10^{-5}$  &  4.22 $\cdot 10^{3}$ \\
  TDHF                       & 118.3  &  0.211 & -1239.4  & 0.081  & 0.995  &  0      		      &  1.22 $\cdot 10^{5}$ \\
  TDRSH                      & 113.3  &  0.171 &  2059.1  & 0.111  & 0.941  &  0      		      &  5.23 $\cdot 10^{4}$ \\
  TDLRSH                     & 114.8  &  0.079 & -1797.2  & 0.087  & 1.000  &  -7.33 $\cdot 10^{-7}$  &  2.78 $\cdot 10^{5}$ \\
  Experiment$^a$  	     & 115.5  &    &    &            \\
\\
 resonance 1s$\to$3p\\[0.1cm]
  TDHF                        &  126.4  & 0.022 & -1279.4 & 0.069  & 1.000 &  5.77  $\cdot 10^{-7}$ & 1.14  $\cdot 10^{5}$  \\
  TDRSH                       &  121.3  & 0.052 &  802.7  & 0.071  & 1.000 & -1.35  $\cdot 10^{-6}$ & 4.60  $\cdot 10^{4}$  \\
  TDLRSH                      &  121.4  & 0.011 & -1791.6 & 0.076  & 1.000 & -2.15  $\cdot 10^{-8}$ & 2.43  $\cdot 10^{5}$  \\  
                        Experiment$^a$   &  121.4  &    &    &          \\
  \hline\hline
\multicolumn{5}{l}{$^a$From Ref.~\onlinecite{CalFleKraMeuPanSta-PRA-90}.}
\end{tabular}
\end{table*}

\begin{figure*}
\includegraphics[scale=0.22,angle=-90]{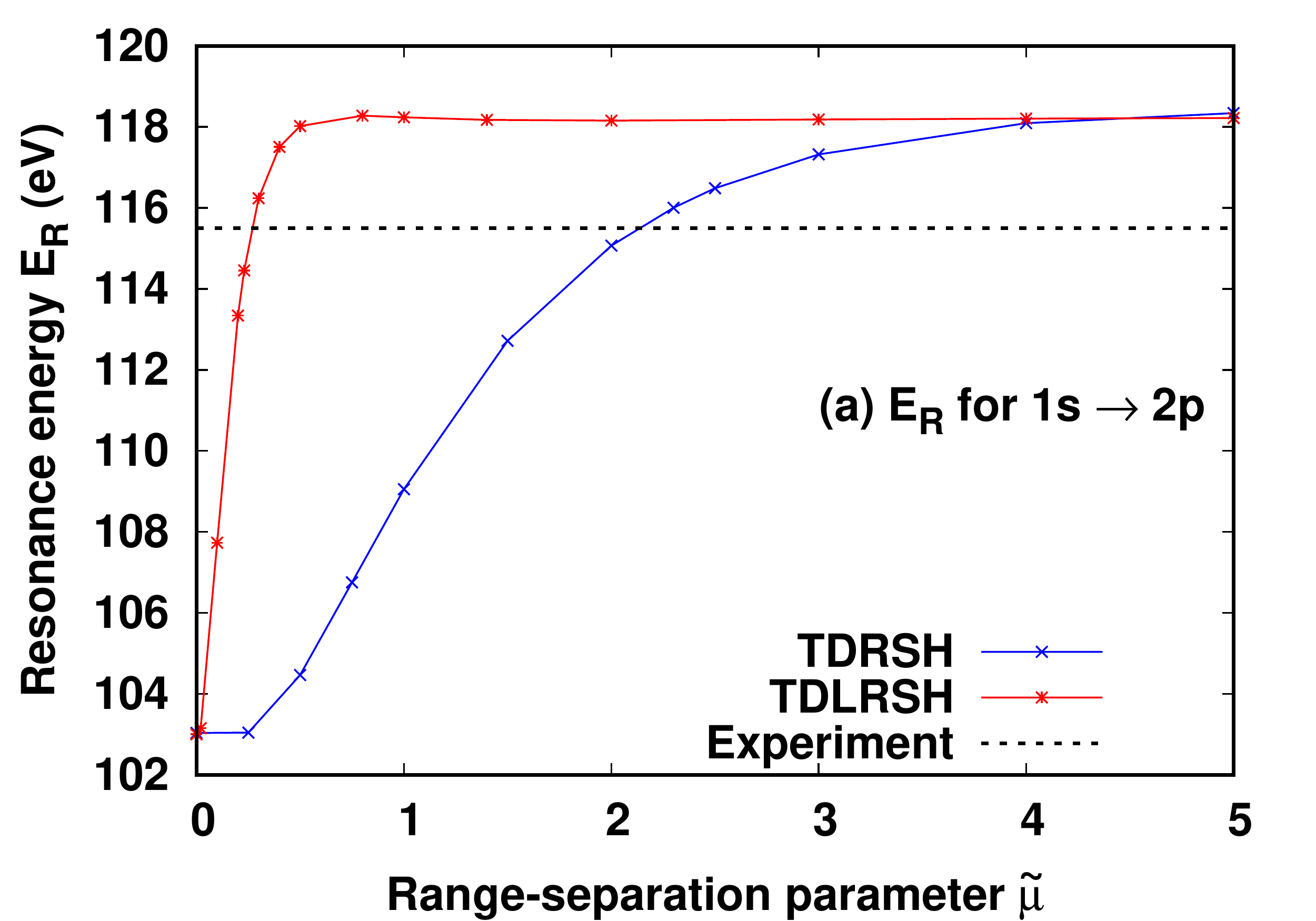}
\includegraphics[scale=0.22,angle=-90]{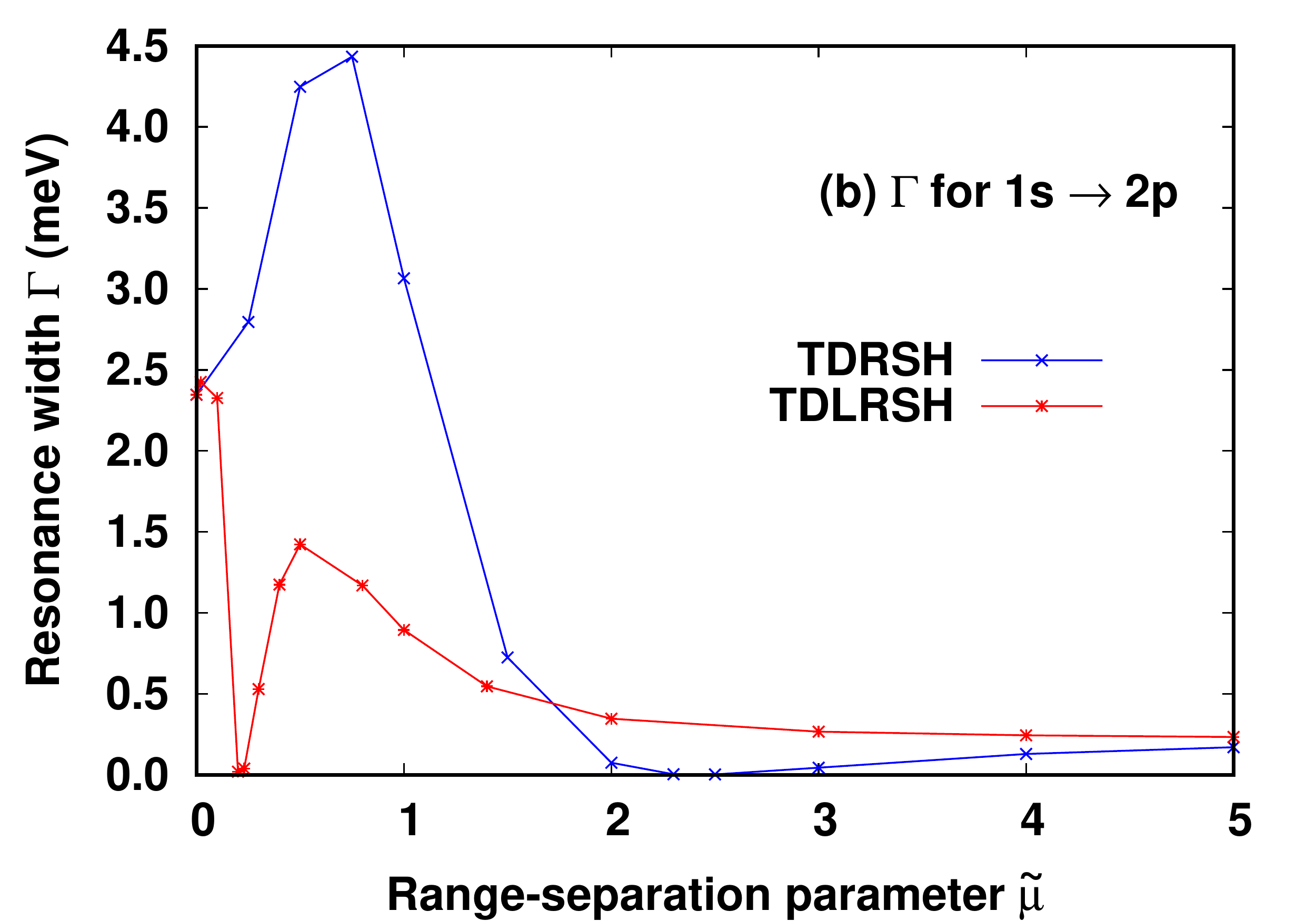}
\includegraphics[scale=0.22,angle=-90]{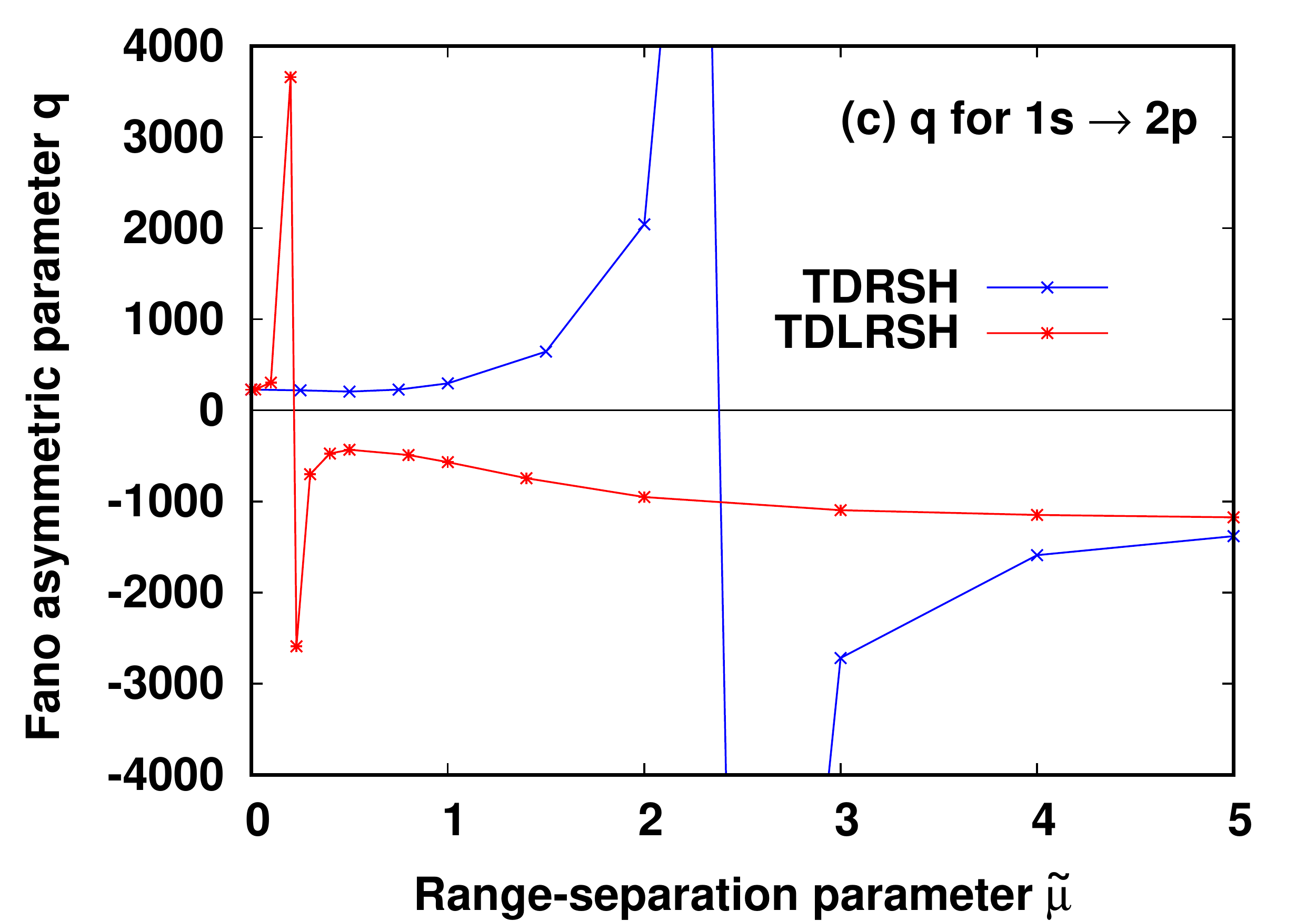}
\caption{Resonance energy $E_\text{R}$, width $\Gamma$, and Fano asymmetric parameter $q$ for the 1s$\to$2p core resonance of the Be atom calculated by TDRSH and TDLRSH as a function of the adimensional range-separation parameter $\tilde{\mu}$.}
\label{fig:parameters}
\end{figure*}

The core resonances (1s$\to$2p, 1s$\to$3p, etc...) are sharp Fesh\-bach resonances, which can be understood as bound states embedded in a continuum turned into quasi-bound states with finite lifetimes due to electron-electron interactions. Figure~\ref{fig:resonance} shows the first 1s$\to$2p resonance obtained with TDLDA, TDHF, TDRSH, and TDLRSH. In all cases, the cross section follows a characteristic asymmetric Fano lineshape which can be fitted to the analytical expression~\cite{FanCoo-PR-65,SteDecLis-JPB-95}
\begin{equation}
\sigma = \sigma_0 (1+a \epsilon) \left[ \rho^2 \frac{(q+\epsilon)^2}{1+\epsilon^2} -\rho^2+1\right],
\label{sigmafit}
\end{equation}
where
\begin{equation}
\epsilon = \frac{\omega-E_\text{R}}{\Gamma/2}.
\end{equation}
Here, $E_\text{R}$ is the resonance energy, $\Gamma$ is the resonance width (or inverse lifetime), $q$ is the asymmetry Fano parameter, $\sigma_0$ is the total background cross section, $a$ is a coefficient for the total background linear drift, and $\rho^2$ is the ratio between the background cross section for transitions to continuum states that interact with the discrete resonant state and the total background cross section. The fitted parameters for the 1s$\to$2p and 1s$\to$3p resonances are given in Table~\ref{tab:resonance}. 
For the fitting procedure, the cross section at the resonance energy $\sigma(E_\text{R})$ was included in the data as the asymmetry parameter $q$ is very sensitive to the value of the cross section at the peak. 
Furthermore, we have verified that, for each resonance, the complex energy $\omega_\text{R} = E_\text{R} -\i \Gamma/2$ is a solution of the non-Hermitian linear-response equation [Eq.~(\ref{Sternheimermatrix})] without external perturbation
\begin{eqnarray}
\left( \begin{array}{cc}
\bm{\Lambda}(\omega_\text{R}) & \b{B} \\
\b{B}^* & \b{\Lambda}(-\omega_\text{R})^*
\end{array} \right)
\left( \begin{array}{c}
\b{c}^{(+)}_\text{R}  \\
\;\b{c}^{(-)*}_\text{R} \\
\end{array} \right)
=\left( \begin{array}{c}
\b{0}  \\
\b{0} \\
\end{array} \right),
\label{Sternheimermatrixresonance}
\end{eqnarray}
with the $\omega_\text{R}$-dependent kinetic integrals in Eq.~(\ref{timunu}) which impose the boundary condition in Eq.~(\ref{Ripmr}). For complex $\omega_\text{R}$ with $\text{Im}[\omega_\text{R}]<0$, this is indeed a Siegert-type boundary condition selecting resonant states that exponentially diverge at infinite distance (see, e.g., Refs.~\onlinecite{Moi-BOOK-11,Hat-JPCS-21}).

The TDLDA 1s$\to$2p resonance occurs at a much too low energy (by 12.5 eV), stemming from the fact that the LDA 1s orbital energy is at a too high energy, as discussed in Sec.~\ref{sec:orbenergies}. The TDHF 1s$\to$2p and 1s$\to$3p resonances occur at too high energies (by 2.8 eV and 5.0 eV), which is consistent with the fact that the HF 1s orbital energy is at a too low energy. The positions of the TDHF resonances (118.3 and 126.4 eV) turn out to be in almost perfect agreement with the values obtained by TDHF calculations using a fairly small Slater basis set~\cite{SteWatDal-JCP-75}, showing that the determination of only the positions of these resonances does not in fact require a large basis set capable of describing continuum states. For comparison, we point out that linear-response time-dependent exact exchange (TDEXX) gives much too low 1s$\to$2p and 1s$\to$3p resonance positions, estimated at about 109 and 111 eV, respectively~\cite{HelBar-JCP-09} (see also Ref.~\onlinecite{SteDecGor-JCP-01} for EXX results). Both TDRSH and TDLRSH give more accurate 1s$\to$2p and 1s$\to$3p resonance positions than TDLDA and TDHF, slightly underestimated by 2.2 eV and 0.1 eV for TDRSH and by 0.7 eV and 0.03 eV for TDLRSH.

The resonance widths $\Gamma$ and Fano asymmetric parameters $q$, which determined the shape of the resonances, are very sensitive to the method employed. TDLDA gives a 1s$\to$2p resonance with the largest width $\Gamma$ and a positive Fano parameter $q$. We note that a similar resonance shape is also obtained when employing more accurate asymptotically corrected exchange-correlation potentials~\cite{SteAltFroDec-CP-97,SteDecGor-JCP-01} or the EXX potential and its adiabatic kernel~\cite{SteDecGor-JCP-01,HelBar-JCP-09}. TDHF gives a sharper 1s$\to$2p resonance with a width $\Gamma$ an order of magnitude smaller and a large negative Fano parameter $q$. The TDHF 1s$\to$3p resonance is even sharper. We note that the shape of the TDHF resonances that we obtain are in agreement with previous relativistic TDHF calculations~\cite{LeeJoh-PRA-80}. TDRSH gives quite sharp resonances with large positive Fano parameters $q$. TDLRSH gives even sharper resonances with large negative Fano parameters $q$. Unfortunately, we have not found accurate resonance widths $\Gamma$ and Fano parameters $q$ in the literature for comparison.

Finally, Figure~\ref{fig:parameters} shows how the resonance energy $E_\text{R}$, width $\Gamma$, and Fano asymmetry parameter $q$ of the 1s$\to$2p resonance calculated by TDRSH and TDLRSH vary with the adimensional range-separation parameter $\tilde{\mu}$, going from the TDLDA limit ($\tilde{\mu}=0$) to the TDHF limit ($\tilde{\mu}\to\infty$). For both TDRSH and TDLRSH the resonance energy increases with $\tilde{\mu}$, until it saturates at the TDHF value. For both TDRSH and TDLRSH, the resonance width does not vary monotonically with $\tilde{\mu}$. In particular, there is a value of $\tilde{\mu}$ (around $\tilde{\mu} \approx 2.5$ and $0.2$ for TDRSH and TDLRSH, respectively) for which the resonance width vanishes. Within the Fano model analysis~\cite{Fan-PR-61}, it means that the coupling between the discrete state corresponding to the resonance and the continuum states vanishes. In this case, the resonance state becomes a truly bound state (with infinite lifetime) embedded in the continuum. At the same critical value of $\tilde{\mu}$, the Fano parameter $q$ jumps from a large positive value to a large negative value. Again, within the Fano model analysis, this may be interpreted as a change of sign of the coupling between the discrete state and the continuum states.

\section{Conclusion}
\label{sec:conclusion}

In this work, we have continued the systematic exploration of linear-response range-separated TDDFT for the calculation of photoionization spectra. We have considered two variants of range-separated TDDFT, namely TDRSH which uses a global range-separation parameter and TDLRSH which uses a local range-separation parameter, and compared with standard TDLDA and TDHF. We have shown how to calculate photoionization spectra with these methods using the Sternheimer approach formulated in a non-orthogonal B-spline basis set. We have illustrated these methods on the photoionization spectrum of the Be atom, focusing in particular on the core resonances.

When the range-separation parameter is adjusted on the 1s ionization edge, both the TDRSH and TDLRSH photoionization spectra constitute a large improvement over the TDLDA photoionization spectrum and a more modest improvement over the TDHF photoionization spectrum. In particular, TDRSH and TDLRSH improve the accuracy of core resonance energies, with a slightly greater accuracy in favor of TDLRSH. 

Possible continuations of this work include testing TDRSH and TDLRSH, as well as possibly other range-separated TDDFT variants, for the calculation of photoionization spectra of more atomic and molecular systems.

\section*{Acknowledgements}
This project has received funding from the CNRS Emergence@INC2021 program (project OPTLHYB) and from the European Research Council (ERC) under the European Union's Horizon 2020 research and innovation programme Grant agreement No. 810367 (EMC2).

\section*{Author Declarations}
The authors have no conflicts to disclose.

\section*{Data Availability}
The data that support the findings of this study are available from the corresponding author upon reasonable request.

\appendix

\section*{Appendix: Boundary conditions for atoms}

In this Appendix, we explain how we impose appropriate boundary conditions at $r=r_\text{max}$ for atoms. 

We first need to study the large-$r$ asymptotic behaviors of the radial solutions $R_{i,\ell}^{(\pm)}$ [Eq.~(\ref{psijrmratoms})] of the Sternheimer equations. For sufficiently large $r$, $R_{i,\ell}^{(\pm)}$ behaves as
\begin{eqnarray}
R_{i,\ell}^{(\pm)}(r,\omega) \isEquivTo{r\to\infty} \bar{R}_{i,\ell}^{(\pm)}(r,\omega),
\label{}
\end{eqnarray}
where $\bar{R}_{i,\ell}^{(\pm)}$ are radial asymptotic solutions of the Sternheimer equations [Eq.~(\ref{Sternheimerrealspace})] which satisfy hydrogen-like Schr\"odinger equations
\begin{eqnarray}
\left( -\frac{1}{2} \frac{\d^2}{\d r^2} + \frac{\ell(\ell+1)}{2r^2}- \frac{Z_\text{eff}}{r} \right) \bar{R}_{i,\ell}^{(\pm)}(r,\omega) \;\;\;\;\;\;\;\;\;\;
\nonumber\\
= (\varepsilon_i \pm \omega +\i \eta) \bar{R}_{i,\ell}^{(\pm)}(r,\omega),
\label{Schroasympt}
\end{eqnarray}
with effective charge $Z_\text{eff}=Z-N+\zeta$. In the expression of $Z_\text{eff}$, the nucleus charge $Z$ and the electron number $N$ come of course from the nucleus-electron potential $v_\ne$ and the Hartree potential $v_\H$, respectively, in the RSH Hamiltonian $h[\gamma_0]$. The contribution $\zeta$ comes from the long-range HF exchange kernel $f_\x^{\lr,\HF}$ and we have $\zeta=1$ for a non-zero range-separation parameter, i.e. $\tilde{\mu} \neq 0$ for the RSH/LRSH scheme. In the LDA limit ($\tilde{\mu}=0$), we have $\zeta=0$. The general solution of Eq.~(\ref{Schroasympt}) may be written as, for $\omega \ge 0$,
\begin{eqnarray}
\bar{R}_{i,\ell}^{(\pm)}(r,\omega) &=& c_1 f_{i,\ell}(r,\pm \omega) + c_2 g_{i,\ell}(r,\pm\omega),
\label{Ripmrgeneral}
\end{eqnarray}
where $c_1$ and $c_2$ are two arbitrary complex-valued coefficients, and the functions $f_{i,\ell}$ and $g_{i,\ell}$ are defined as, for a general frequency variable $z\in \mathbb{C}$,
\begin{eqnarray}
f_{i,\ell}(r,z) = F_{\ell}(- Z_\text{eff}/k_i(z),k_i(z) r),
\label{}
\end{eqnarray}
\begin{eqnarray}
g_{i,\ell}(r,z) = G_{\ell}(-Z_\text{eff}/k_i(z),k_i(z) r),
\label{}
\end{eqnarray}
where $F_{\ell}$ and $G_{\ell}$ are the regular and irregular Coulomb functions~\cite{AbrSte-BOOK-83}, and $k_i(z)=\sqrt{2(\varepsilon_i + z +\i \eta)}$ is the complex-valued momentum.
The asymptotic behavior of the Coulomb functions are
\begin{eqnarray}
  \label{eq_fi}
f_{i,\ell}(r,z) \isEquivTo{r\to\infty} \sin \theta_{i,\ell}(r,z),
\end{eqnarray}
\begin{eqnarray}
  \label{eq_gi}
g_{i,\ell}(r,z) \isEquivTo{r\to\infty} \cos\theta_{i,\ell}(r,z),
\end{eqnarray}
with 
\begin{eqnarray}
  \label{eq:thetai}
\theta_{i,\ell} (r,z) = k_i(z) r + \frac{Z_\text{eff}}{k_i(z)} \ln(2 k_i(z) r) - \frac{1}{2} \ell \pi + \sigma_{i,\ell}(z),
\end{eqnarray}
and $\sigma_{i,\ell}(z) = \arg \Gamma ( \ell +1 - \i  Z_\text{eff}/k_i(z) )$. As $r\to\infty$, $\theta_{i,\ell} (r,\pm \omega) \sim k_i(\pm \omega) r$ and due to the fact that $\text{Im}[k_i(\pm\omega)]>0$ (for $\eta>0$), we see that the general solution $\bar{R}_{i,\ell}^{(\pm)}$ in Eq.~(\ref{Ripmrgeneral}) $\bar{R}_{i,\ell}^{(\pm)}$ does not diverge as $r\to\infty$ only for the coefficient ratio $c_1/c_2=\i$, so we have
\begin{eqnarray}
\bar{R}_{i,\ell}^{(\pm)}(r,\omega) &=& c_2 \left[ \i f_{i,\ell}(r,\pm\omega) + g_{i,\ell}(r,\pm\omega) \right],
\label{Ripmr}
\end{eqnarray}
and then $\bar{R}_{i,\ell}^{(\pm)}(r,\omega)\isEquivTo{r\to\infty} c_2 \exp (\i \, \theta_{i,\ell}(r,\pm\omega))$.

We can thus impose a Robin boundary condition for $R_{i,\ell}^{(\pm)}$ at $r=r_\text{max}$ of the form
\begin{eqnarray}
\left.\frac{\d \ln R_{i,\ell}^{(\pm)}(r,\omega)}{\d r}\right|_{r=r_\text{max}} =  b_{i,\ell}(\pm\omega),
\label{boundarycondrmax}
\end{eqnarray}
with $b_{i,\ell}(\pm\omega)=\d \ln \bar{R}_{i,\ell}^{(\pm)}(r,\omega)/\d r |_{r=r_\text{max}}$. Even though this expression of $b_{i,\ell}(\pm\omega)$ works for both $R_{i,\ell}^{(+)}$ and $R_{i,\ell}^{(-)}$ and for any $\omega$, it is insightful to further look at the physical content of the asymptotic behavior of $R_{i,\ell}^{(+)}$ and $R_{i,\ell}^{(-)}$, separately, in the limit $\eta \to 0^+$. As regards $R_{i,\ell}^{(+)}$, for $\omega \geq -\varepsilon_i$, we have $k_i(\omega)=\text{Re}[k_i(\omega)]=\sqrt{2(\varepsilon_i + \omega)}$, and the asymptotic solution $\bar{R}_{i,\ell}^{(+)}$ is an outgoing spherical wave. In this case, the explicit expression of $b_{i,\ell}(\omega)$ that we use is 
\begin{eqnarray}
b_{i,\ell}(\omega) = \left. \frac{\i \; \d f_{i,\ell}(r,\omega)/\d r + \d g_{i,\ell}(r,\omega)/\d r}{\i f_{i,\ell}(r,\omega) + g_{i,\ell}(r,\omega)} \right|_{r=r_\text{max}},
\nonumber\\
\text{if $\omega \geq-\varepsilon_i$}. \;\;\;\;\;\;\;\;\;
\label{blargeomega}
\end{eqnarray}
For $\omega <-\varepsilon_i$, we have $k_i(\omega)=\i \; \text{Im}[k_i(\omega)] = \i \sqrt{-2(\varepsilon_i + \omega)}$, and the asymptotic solution $\bar{R}_{i,\ell}^{(+)}$ decays exponentially. Instead of using Eq.~(\ref{blargeomega}), we can more simply use
\begin{eqnarray}
b_{i,\ell}(\omega) = 0, \; \text{if $\omega <-\varepsilon_i$},
\label{bsmallomega}
\end{eqnarray}
which corresponds to imposing Neumann boundary condition at $r=r_\text{max}$. Similarly, for $R_{i,\ell}^{(-)}$, since we always have $\varepsilon_i - \omega <0$ (because $\varepsilon_i < 0$ for occupied orbitals), we have $k_i(-\omega)=\i \; \text{Im}[k_i(-\omega)] = \i \sqrt{-2(\varepsilon_i - \omega)}$, thus the asymptotic solution $\bar{R}_{i,\ell}^{(-)}$ decays exponentially, and we can use again
\begin{eqnarray}
b_{i,\ell}(-\omega) = 0.
\label{bmomega}
\end{eqnarray}

\medskip

\begin{figure}[t]
\includegraphics[scale=0.3,angle=-90]{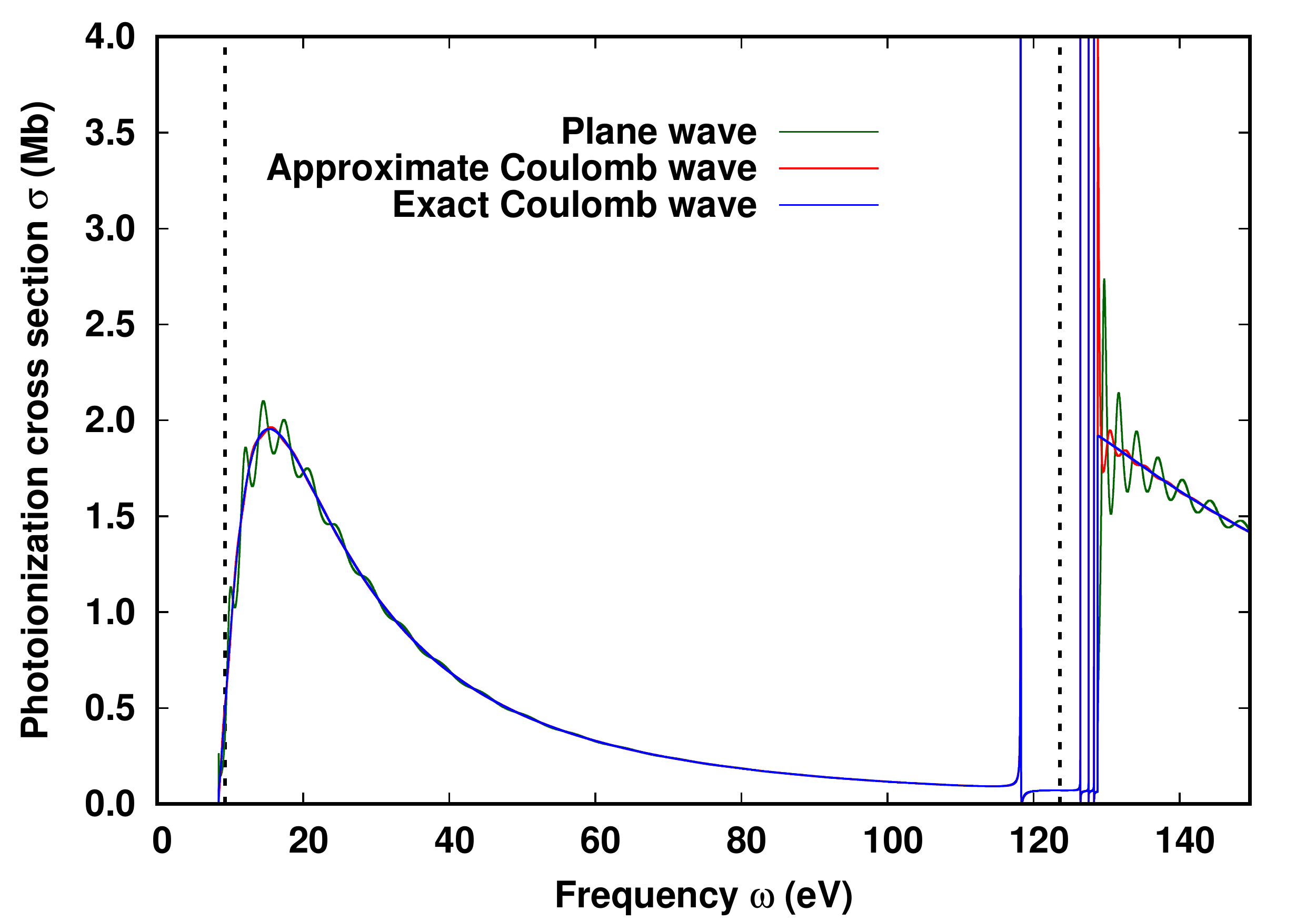}
\caption{Photoionization cross section of the Be atom calculated with TDHF using different boundary conditions: plane wave [Eq.~(\ref{PW})], approximate Coulomb wave [Eq.~(\ref{ACW})], and exact Coulomb wave [Eq.~(\ref{Ripmr})].}
\label{fig:boundary}
\end{figure}

\medskip

For a neutral system such as like the Be atom, we have $Z_\text{eff}=1$ for the TDHF method
and using the exact Coulomb-wave boundary condition in Eq.~(\ref{Ripmr}) for $\omega \geq-\varepsilon_i$
proves crucial to obtain converged results with a modest maximal radius $r_{\text{max}}$.
To show this, we compare in Figure~\ref{fig:boundary} two alternative
types of boundary conditions for the same maximal radius of $r_{\text{max}} = 25$ bohr. 
First, we can keep the correct
long-range behavior but make the asymptotic approximation
in Eqs.~(\ref{eq_fi})-(\ref{eq:thetai}), valid for $k_i(\omega)r \gg 1$, resulting in
the approximate Coulomb-wave (ACW) boundary condition
\begin{eqnarray}
\bar{R}_{i,\text{ACW}}^{(+)}(r,\omega) = c_2 \exp \left[ \i \left( k_i(\omega)r + \frac{Z_\text{eff}}{k_i(\omega)} \ln \left( 2 k_i(\omega) r \right) \right) \right].
\nonumber\\
\label{ACW}
\end{eqnarray}
At the scale of the plot, the part of the spectrum corresponding to the 2s ionization is barely affected, but this creates oscillations beyond the 1s ionization edge.
A more drastic approximation is to further ignore the Coulomb potential, i.e.
setting $Z_\text{eff}=0$. In this case, we get the plane-wave (PW) boundary
condition
\begin{eqnarray}
\bar{R}_{i,\text{PW}}^{(+)}(r,\omega) &=& c_2 \exp \left( \i k_i(\omega)r \right).
\label{PW}
\end{eqnarray}
This creates strong and rapid oscillations near both ionization thresholds.
However, far from the ionization thresholds, the effect is less pronounced. In
particular, the 1s$\to$2p resonance is still well reproduced. This is
because this resonance results from the interaction of a bound state
with continuum states at a relatively large energy,
which behave almost like plane waves. This can be understood from
Eq.~(\ref{eq:thetai}), which implies that the plane-wave approximation is
reasonable when
\begin{align}
 k_{i}(\omega)^{2} r_{\text{max}} \gg  {Z_{\text{eff}}} \ln(2 k_{i}(\omega) r_{\text{max}}),
\end{align}
i.e. when the wavelength of the outgoing
wave is small compared to $\sqrt{{r_{\text{max}}}/{Z_{\text{eff}}}}$.
In both cases, these oscillations can be reduced by either increasing the computational box size $r_{\text{max}}$ or by using a non-zero broadening factor $\eta$. The former increases the computational effort while the latter may lead to a loss of features in the spectrum. The use of the exact Coulomb-wave boundary condition removes these oscillations without the need of increasing $r_{\text{max}}$ or using a non-zero $\eta$.

In the case of TDLDA, we use the boundary condition in Eq.~(\ref{Ripmr}) with $Z_\text{eff}=0$, giving a free spherical outgoing wave (the Coulomb functions reduce to the Riccati-Bessel functions).
This is a better boundary condition than using the plane-wave approximation in Eq.~(\ref{PW}), the latter giving slight oscillations for TDLDA.

Finally, we use the boundary condition in Eq.~(\ref{Ripmr}) with $Z_\text{eff}=\erf(\mu r_\text{max})$ for TDRSH and with $Z_\text{eff}=\erf(\mu(r_\text{max}) r_\text{max})$ for TDLRSH, which goes smoothly from $Z_\text{eff}=0$ for $\mu \ll 1/r_\text{max}$ or $\mu (r_\text{max}) \ll 1/r_\text{max}$ to $Z_\text{eff}=1$ for $\mu \gg 1/r_\text{max}$ or $\mu (r_\text{max}) \gg 1/r_\text{max}$.


\end{document}